\documentclass[preprint,pre,twocolumn]{revtex4}%
\usepackage{amssymb}
\usepackage{amsfonts}
\usepackage{amsmath}
\usepackage{graphicx}%
\providecommand{\U}[1]{\protect\rule{.1in}{.1in}}

\begin{document}
\preprint{ }
\title[ ]{Nonlinear higher-order hydrodynamics. Unification of kinetic and hydrodynamic
approaches within a nonequilibrium statistical ensemble formalism.}
\author{C. A. B. Silva}
\affiliation{\textit{Departamento de F\'{\i}sica, Instituto Tecnol\'{o}gico de
Aeron\'{a}utica, 12228-901, S\~{a}o Jos\'{e} dos Campos, SP, Brazil.}}
\author{J. Galv\~{a}o Ramos, Aurea R. Vasconcellos and Roberto Luzzi *\thanks{*Group
Home Page: \ www.ifi.unicamp.br/$\sim$aurea}}
\affiliation{Condensed Matter Physics Department, \textit{Institute of Physics "Gleb
Wataghin\textquotedblright, State University of Campinas - Unicamp, 13083-859,
Campinas, SP, Brazil.}}

\begin{abstract}
Construction, in the framework of a Nonequilibrium Statistical Ensemble
Formalism, of a Mesoscopic Hydro-Thermodynamics, that is, covering phenomena
involving motion displaying variations short in space and fast in time
--unrestricted values of Knudsen numbers--, is presented. In that way, it is
provided an approach enabling for the coupling and simultaneous treatment of
the kinetics and hydrodynamic levels of descriptions. \ It is based on a
complete thermo-statistical approach in terms of the densities of matter and
energy and their fluxes of all orders covering systems arbitrarily driven away
from equilibrium. The set of coupled nonlinear integro-differential
hydrodynamic equations is derived. They are the evolution equations of the
Grad-like moments of all orders, derived from a generalized kinetic equation
built in the framework of the Nonequilibrium Statistical Ensemble Formalism.
For illustration, the case of a system of particles embedded in a fluid acting
as a thermal bath is fully described. The resulting enormous set of coupled
evolution equations is of unmanageable proportions, thus requiring in practice
to introduce an appropriate description using the smallest possible number of
variables. We have obtained a hierarchy of Maxwell times, which can be
considered a kind of Bogoliubov's characteristic times in hydrodynamic and
which have a particular relevance in the criteria for stablishing a
contraction of description.

\end{abstract}
\maketitle

\section{INTRODUCTION}

It has been noticed that one of the complicated problems of the nonequilibrium
theory of transport processes in dense gases and liquids is the fact that
their kinetics and hydrodynamics are intimately coupled, and must be treated
simultaneously (e.g., see Refs. \cite{klimontovich}-\cite{zubarev1}). On this
we may say that microscopic descriptions of hydrodynamics, that is, associated
to a derivation of kinetic equations from classical or quantum mechanics and
containing kinetic (transport) coefficients written in terms of correlation
functions, is a long standing traditional problem. An important aspect is the
derivation of constitutive laws which express thermodynamic fluxes (or
currents, as those of matter and energy) in terms of appropriate thermodynamic
forces (typically gradients of densities as those of matter and energy). In
their most general form these laws are nonlocal in space and non-instantaneous
in time.\ A first kinetic-hydrodynamic approach can be considered to be the
so-called \textit{classical (or Onsagerian) hydrodynamics}; it gives
foundations to, for example, the classical Fourier%
\'{}%
s and Fick%
\'{}%
s diffusion laws. But it works under quite restrictive conditions, namely,
local equilibrium; linear relations between fluxes and thermodynamic forces
(meaning weak amplitudes in the motion) \ with Onsager%
\'{}%
s symmetry laws holding; near homogeneous and static movement (meaning that
the motion can be well described with basically Fourier components with long
wavelengths and low frequencies, and then involves only smooth variation in
space and time); weak and rapidly regressing fluctuations \cite{kreuzer}%
-\cite{casimir}.

Hence, more advanced approaches are required to lift these restrictions.
Consider first near homogeneity, which implies validity in the limit of long
wavelengths (or wavenumber $Q$ approaching zero). To go beyond it is necessary
to introduce a proper dependence on $Q$ valid, in principle, for intermediate
and short wavelengths (intermediate to large wavenumbers). In phenomenological
theories this corresponds to go from classical irreversible thermodynamics to
extended irreversible thermodynamics \cite{jou}-\cite{jou1}. This is what has
been called \textit{generalized hydrodynamics}, a question extensively debated
for decades by the Statistical Mechanics community. Several approaches have
been used, and a description can be consulted in Chapter 6 of the classical
book on the subject by Boon and Yip \cite{boon}. Introduction of nonlocal
effects for describing motions with influence of ever decreasing wavelengths,
going towards the very short limit, has been done in terms of expansions in
increasing powers of the wavenumber, which consists in what is sometimes
referred to as \textit{higher-order hydrodynamics} (HOH). Attempts to perform
such expansions are the so-called Burnett and super-Burnett approaches in the
case of mass motion, and Guyer-Krumhansl approach in the case of propagation
of energy (see for example Refs. [11] and [12]). An usual approach has been
based on the moments solution procedure of Boltzmann equation, as in the work
of Hess \cite{hess}, using a higher-order Chapman-Enskog solution method. The
Chapman-Enskog method provides a solution to Boltzmann equation consisting of
a series in powers of the Knudsen number, $K_{n}$, given by the ratio between
the mean-free path of the particles and the scale of change (relevant
wavelengths in the motion) of the hydrodynamic fields. Retaining the term
linear in $K_{n}$ there follows Navier-Stokes equation, the term in $K_{n}%
^{2}$ introduces Burnett-like contributions, and the higher-order ones
($K_{n}^{3}$ and up) the super-Burnett contributions.

A satisfactory development of a HOH being also nonlinear and including
fluctuations is highly desirable for covering a large class of hydrodynamic
situations, and, besides its own scientific interest, also for obtaining
insights into present day technological-industrial processes. Also, we can
mention its fundamental relevance in Oceanography and Meteorology (e.g.
\cite{wunch}, \cite{heeling}), and that it has been stated \cite{chiu} that
the idea of promoting hydraulics by statistical inference is appealing because
the complete information about phenomena in hydraulics seldom exists; for
example sediment transport and the more fundamental problem in fluid mechanics
of describing the velocity distribution in fluids under flow \cite{jou4}%
-\cite{vasconcellos}. Indeed, the nonlocal terms become specially important in
miniaturized devices at submicronic lengths \cite{anile}, or in the design of
stratospheric planes, which fly in rarefied gases in a density regime between
the independent particle description and the purely continuous description.
Another particular problem to it related is the one of obtaining the
structures of shock waves in fluids for wide ranges of Mach numbers
\cite{uribe}. Moreover, Burnett approximation of hydrodynamics has been shown
to provide substantial improvement on many features of the flow occurring in
several problems in hydrodynamics, e.g. the case of Poiseuille flow
\cite{uribe1} and others \cite{mackowski}.

The microscopic derivation of a HOH, together with the analysis of the
validity of existing theories, is still a point in question. It has been shown
\cite{bobylev} that for the case of Maxwellian molecules, whereas
Navier-Stokes approximation yields equations which are stable against small
perturbations, this is not the case when are introduced Burnett contributions
to the equations. It follows that small perturbations to the solutions, which
are periodic in the space variable with a wavelength smaller than a critical
length, are exponentially unstable. This fact has been called
\textit{Bobylev's instability}. Moreover, Karlin \cite{karlin} reconsidered
the question looking for exact solutions to simplified models: When a
linearized ten-moments Grad method is used, and the Chapman-Enskog method is
applied to the model, does in fact there follow instabilities in the
higher-order approximations. On the other hand, resorting to the
Chapman-Enskog solution for linearized Grad ten-moment equations resummed
exactly, solutions are obtained for which the stability of higher-order
hydrodynamics, in various approximations, can be discussed. However, more
recently, Garcia-Colin and collaborators \cite{uribe2} have extended Bobylev's
analysis for the case of any interaction potential, and have demonstrated that
one can interpret the fact as to give a bound for a Knudsen number above which
the Burnett equations are not valid, with no instability involved.

Furthermore, inclusion of nonlinearity in the theory, in a Mesoscopic
Hydro-Thermodynamics (MHT for short and meaning thermal physics of fluid
continua), leads to additional possible singularities, called hydrodynamic
singularities, as, for example, those described in Refs. \cite{eggers}%
-\cite{godreche}. A satisfactory construction of a MHT is highly desirable for
covering a large class of hydrodynamic situations obtaining an understanding
of the physics involved from the microscopic level, and in the last instance
gaining insights into technological and industrial processes as in, for
instance, hydraulic engineering, food engineering, soft-matter engineering,
oil production and petrochemistry, etc., which have an associated economic
interest. MHT was initiated by the so-called Catalan School of Thermodynamics
as a large expansion of Extended Irreversible Thermodynamics
\cite{dedeurwaerdere}.

It can be noticed that nowadays two approaches appear to be the most favorable
for providing very satisfactory methods to deal with hydrodynamics within an
ample scope of nonequilibrium conditions. They are Nonequilibrium \ Molecular
Dynamics (NMD) \cite{adler} and the kinetic theory based on the far reaching
generalization of Gibbs' ensemble formalism, namely the Nonequilibrium
Statistical Ensemble Formalism (NESEF for short) \cite{luzzi}-\cite{akhiezer}.
NMD is a computational method created for modeling physical system at the
microscopic level, being a good technique to study the molecular behavior of
several physical processes. Together with the so-called Monte Carlo method are
part of what is known as numeric simulation methods \cite{kalos}.

We do here present an extensive derivation of a MHT on the basis of the
kinetic theory founded on NESEF, quite appropriate to deal with systems in
far-from-equilibrium conditions involving the development of ultrafast
relaxation processes, and displaying nonlinear behavior leading, eventually,
to instabilities and synergetic self-organization \cite{nicolis}%
-\cite{glandsdorff}. Within the framework of NESEF, but in a different
approach to the one used here, an alternative MHT was introduced by Zubarev
and Tishchenko \cite{zubarev3}, \cite{zubarev4}.

It may be noticed that the formalism can be extended to deal with the
so-called non-conventional hydrodynamics which is associated to disordered
media \cite{hawling}, consisting in systems showing a complex structure of a
fractal-like (self-affine in average) characteristics, whose range of
applicability and of physical interest is large \cite{family}. Fall on this
problem the case of the distinctive behavior of polyatomic structures such as
colloidal particles, surfactant micelles, and polymer and biopolymer (as DNA)
in liquid solutions, which are classical examples of what is presently
referred to as soft condensed matter \cite{witten}. One particular case of
apparently unusual behavior is the one associated to hydrodynamic motion
leading to a so-called non-Fickian diffusion, described by a time evolution
following a kind of fractional-power law \cite{crank}. \ The nonequilibrium
statistical thermo-mechanical aspects of complex systems including
illustrations is reported elsewhere \cite{luzzi6}; a case involving
hydro-thermodynamics is given in Ref. \cite{ramos}.

In the present paper the conventional NESEF-based MHT is described in next
Section, accompanied with the study of a system consisting of particles
embedded in a fluid which acts as a thermal bath at rest and in thermal
equilibrium with an external reservoir. The general theory for the MHT is
built upon a generalization of Grad's moments method for solution of, in this
case, a generalized kinetic equation derived in the context of NESEF
\cite{silva}.

\section{THEORETICAL\ BACKGROUND}

For building a nonlinear higher-order (generalized) hydro-thermodynamics on
mechanical statistical basis, one needs to resort to a nonequilibrium
statistical ensemble formalism (NESEF) for open systems. Such formalism was
developed step by step along the past century by a number of renowned
scientists whose contributions have been systematized and generalized in a
close structure, as described in Refs. \cite{luzzi} to \cite{akhiezer}.

According to theory, immediately after the open system of $N$ particles, in
contact with external sources and reservoirs, has been driven out of
equilibrium, the description of its state requires to introduce all its
observables, their fluctuations and, eventually, higher-order variances. In
most cases it suffices to take a reduced set of observables, what implies in
to have access to the so-called one-particle (or single-particle), $\hat
{n}_{1}$, and two-particle, $\hat{n}_{2}$, dynamical operators for any subset
of the particles involved. This is so because all observable quantities can be
expressed at the microscopic mechanical level in terms of these operators
(e.g. Refs. \cite{fano} and \cite{bogoliubov}).

On the basis of the construction of the nonequilibrium statistical operator
\cite{luzzi}-\cite{luzzi1}, and taking into account the noted above fact that
a complete description of the nonequilibrium state of the system follows from
the knowledge of the single- and two-particle density operators (or
equivalently the associated reduced density matrices) which in classical
mechanics are
\begin{equation}
\widehat{n}_{1}(\mathbf{r},\mathbf{p})=\sum_{j=1}^{N}\delta\left(
\mathbf{r}-\mathbf{r}_{j}\right)  \delta\left(  \mathbf{p}-\mathbf{p}%
_{j}\right)  \text{,}%
\end{equation}%
\[
\widehat{n}_{2}(\mathbf{r},\mathbf{p};\mathbf{r}^{,},\mathbf{p}^{,}%
)=\sum_{j\neq k=1}^{N}\delta\left(  \mathbf{r}-\mathbf{r}_{j}\right)
\delta\left(  \mathbf{p}-\mathbf{p}_{j}\right)
\]%
\begin{equation}
\times\delta\left(  \mathbf{r}^{,}-\mathbf{r}_{k}\right)  \delta\left(
\mathbf{p}^{,}-\mathbf{p}_{k}\right)  \text{,}%
\end{equation}
where $\mathbf{r}_{j}$, $\mathbf{p}_{j}$ are the coordinate and momentum of
the j-th particle, and $\mathbf{r}$, $\mathbf{p}$ are called field variables,
the most complete nonequilibrium statistical distribution \cite{zubarev},
\cite{luzzi}-\cite{akhiezer} is the one built in terms of the auxiliary
statistical operator
\begin{equation}
\bar{{\mathcal{R}}}(t,0)=\bar{\rho}(t,0)\times\rho_{R}, \label{eq3}%
\end{equation}
where $\bar{\rho}$ refers to the system of $N$ particles of mass $m$, and
$\rho_{R}$ is the one associated to a thermal bath of $N_{R}$ particles of
mass $M$ taken in equilibrium at temperature $T_{0}$. The first one is given
by
\[
\bar{\rho}(t,0)=\exp\Bigg\{-\phi(t)-\int d^{3}rd^{3}p\text{ }F_{1}%
({\mathbf{r}},{\mathbf{p}};t)
\]%
\[
\times\hat{n}_{1}({\mathbf{r}},{\mathbf{p}})
\]%
\[
-\int d^{3}rd^{3}p\int d^{3}r^{\prime}d^{3}p^{\prime}\text{ }F_{2}%
({\mathbf{r}},{\mathbf{p}},{\mathbf{r}}^{\prime},{\mathbf{p}}^{\prime};t)
\]%
\begin{equation}
\times\,\hat{n}_{2}({\mathbf{r}},{\mathbf{p}},{\mathbf{r}}^{\prime
},{\mathbf{p}}^{\prime})\Bigg\}. \label{eq4}%
\end{equation}
Hence, $\bar{\rho}(t,0)$ depends on the variables of the system of interest
and $\rho_{R}$ on the variables of the thermal bath; both distributions are
taken as normalized, as it should, with $\phi(t)$ ensuring the normalization
of $\bar{\rho}$, that is,%

\[
\phi(t)=\int d\Gamma\,\exp\Bigg\{-\int d^{3}rd^{3}p\text{ }F_{1}({\mathbf{r}%
},{\mathbf{p}};t)\,
\]%
\[
\times\hat{n}_{1}({\mathbf{r}},{\mathbf{p}})
\]%
\[
-\int d^{3}rd^{3}p\int d^{3}r^{\prime}d^{3}p^{\prime}\text{ }F_{2}%
({\mathbf{r}},{\mathbf{p}},{\mathbf{r}}^{\prime},{\mathbf{p}}^{\prime};t)
\]%
\begin{equation}
\times\,\hat{n}_{2}({\mathbf{r}},{\mathbf{p}},{\mathbf{r}}^{\prime
},{\mathbf{p}}^{\prime})\Bigg\},
\end{equation}
and $F_{1}$ and $F_{2}$ are the nonequilibrium thermodynamic variables
conjugated to $\hat{n}_{1}$ and $\hat{n}_{2}$ meaning that%

\begin{equation}
\frac{\delta\ln\overline{Z}(t)}{\delta F_{1}\left(  \mathbf{r},\mathbf{p}%
;t\right)  }=-Tr\left\{  \widehat{n}_{1}(\mathbf{r},\mathbf{p})\bar{\rho
}(t,0)\right\}  ,
\end{equation}%
\begin{equation}
\frac{\delta\ln\overline{Z}(t)}{\delta F_{2}\left(  \mathbf{r},\mathbf{p}%
,\mathbf{r}^{\prime},\mathbf{p}^{\prime};t\right)  }=-Tr\left\{
\widehat{n}_{2}(\mathbf{r},\mathbf{p};\mathbf{r}^{,},\mathbf{p}^{,})\bar{\rho
}(t,0)\right\}  ,
\end{equation}
where ln$\overline{Z}(t)=\phi(t)$ with $\overline{Z}(t)$ playing the role of a
nonequilibrium partition function and $\delta$ stands for functional
derivative, in complete analogy with the equilibrium case. Moreover, $d\Gamma$
is the element of volume in the phase space of the system, and for simplicity
we have omitted to indicate the dependence on $\Gamma$ of $\hat{n}_{1}$,
$\hat{n}_{2}$, $\bar{\rho}$, $\bar{{\mathcal{R}}}$, and that $\rho_{R}$
depends on the phase point $\Gamma_{R}$ in the phase space of the bath.

We stress that $\bar{\rho}$ of Eq.(\ref{eq4}) is not the statistical operator
of the nonequilibrium system, but an auxiliary one -- called the
\ \textquotedblleft instantaneously frozen quasi-equilibrium\textquotedblright%
\ statistical operator -- that allows to built the proper nonequilibrium
statistical operator, which needs to include \textit{\ historicity and
irreversibility} not present in $\bar{\rho}$, hence it does not account for
dissipative processes, and besides does not provide correct average values in
the calculation of transport coefficients and response functions.

We recall that the nonequilibrium statistical operator is given by
\cite{zubarev}, \cite{luzzi}-\cite{akhiezer}
\begin{align}
{\mathcal{R}}_{\varepsilon}(t)  &  =\exp\Bigg\{\ln\bar{\rho}(t,0)-\int%
_{-\infty}^{t}dt^{\prime}e^{\varepsilon(t^{\prime}-t)}\nonumber\\
&  \frac{d\hfill}{dt^{\prime}}\ln\bar{\rho}(t^{\prime},t^{\prime
}-t)\Bigg\}\times\rho_{R}, \label{eq811}%
\end{align}
with $\bar{\rho}(t,0)$ of Eq. (\ref{eq4}), and where%
\begin{equation}
\bar{\rho}(t^{\prime},t^{\prime}-t)=\exp\Big\{i(t-t^{\prime}){\mathcal{L}%
}\Big\}\bar{\rho}(t^{\prime},0), \label{eq91}%
\end{equation}
is the auxiliary operator carrying on the mechanical evolution of the system
under Hamiltonian $\hat{H}$ (${\mathcal{L}}$ is the Liouvillian operator of
the system meaning $i{\mathcal{L}}\hat{A}=\{\hat{A},\hat{H}\}$). Usually the
system's Hamiltonian is separated out into two terms, namely,%

\begin{equation}
\hat{H}=\hat{H}_{0}+\hat{H}^{\prime}\text{,} \label{eq6}%
\end{equation}
where $\hat{H}_{0}$ is the kinetic energy operator and%

\begin{equation}
\hat{H}^{\prime}=\hat{H}_{1}+\hat{W}+\hat{H}_{P}\text{,} \label{eq7}%
\end{equation}
contains the internal interactions energy operator $\hat{H}_{1}$, while
$\hat{W}$ accounts for the interaction of the system with the thermal bath,
and $\hat{H}_{P}$ is the energy operator associated to the coupling of the
system with external pumping sources. Finally, $\varepsilon$ is an
infinitesimal positive real number which is taken going to zero after the
traces in the calculation of averages have been performed (it is present in a
kernel that introduces irreversibility in the calculations, in a
Krylov-Bogoliubov sense \cite{luzzi}-\cite{zubarev2}). We stress that the
second contribution in the exponent in Eq. (\ref{eq811}) accounts for
historicity and irreversible evolution from the initial time (taken in the
remote past, $t_{0}\rightarrow-\infty$, implying in adiabatic coupling of
correlations, (see for example Ref. \cite{luzzi}), or alternatively, can be
seen as the adiabatic coupling of the interactions responsible for relaxation
processes \cite{alvarez}). Moreover we notice that the time derivative in Eq.
(\ref{eq811}) takes care of the change in time of the thermodynamic state of
the system (the first term in the argument i.e. $t^{\prime}$) and of the
microscopic mechanical evolution (second term in the argument, i.e.
$t^{\prime}-t$, see Eq. (\ref{eq91})), and that the initial value condition is
${\mathcal{R}}_{\varepsilon}(t_{0})=\bar{\rho}(t_{0},0)$ for $t_{0}%
\rightarrow-\infty$.

The nonequilibrium thermodynamic space of states \cite{luzzi2} associated to
the basic dynamic variables $\hat{n}_{1}$ and $\hat{n}_{2}$\ is composed by
the one-particle and two-particle distribution functions
\begin{align}
f_{1}\left(  \mathbf{r},\mathbf{p};t\right)   &  =Tr\left\{  \widehat{n}%
_{1}\left(  \mathbf{r},\mathbf{p}\right)  \varrho_{\varepsilon}\left(
t\right)  \right\} \nonumber\\
&  =Tr\left\{  \widehat{n}_{1}\left(  \mathbf{r},\mathbf{p}\right)
\overline{\varrho}\left(  t,0\right)  \right\}  \label{eq8}%
\end{align}%
\begin{align}
f_{2}\left(  \mathbf{r},\mathbf{p},\mathbf{r}^{,},\mathbf{p}^{,};t\right)   &
=Tr\left\{  \widehat{n}_{2}\left(  \mathbf{r},\mathbf{p},\mathbf{r}%
^{,},\mathbf{p}^{,}\right)  \varrho_{\varepsilon}\left(  t\right)  \right\}
\nonumber\\
&  =Tr\left\{  \widehat{n}_{2}\left(  \mathbf{r},\mathbf{p},\mathbf{r}%
^{,},\mathbf{p}^{,}\right)  \overline{\varrho}\left(  t,0\right)  \right\}
\label{eq9}%
\end{align}
where we indicate that for the basic variables, and only for them, the average
with the statistical operator $\varrho_{\varepsilon}$ is equal to the one
taken with the auxiliary operator \cite{luzzi}-\cite{zubarev2}. The trace
operation $Tr$ is in this classical approach to be understood as an
integration over phase space; $\hat{n}_{1}$ and $\hat{n}_{2}$ are functions on
phase space and $\overline{\varrho}$ and $\varrho_{\varepsilon}$ functionals
of these two. The knowledge of the two distribution functions $f_{1}$ and
$f_{2}$ allows to determine the value and evolution of any observable of the
system as well as of response functions and transport coefficients.

The knowledge of $f_{1}\left(  \mathbf{r},\mathbf{p};t\right)  $ implies
complete information about the actual distribution of particles, and therefore
of the physical properties of the system. Alternatively, knowing all the
moments of the distribution allows to have a complete knowledge of its
characteristics. A knowledge of some moments is not sufficient to determine
the distribution completely; it implies in only possessing partial knowledge
of the characteristics of this distribution \cite{reif}. Grad noticed that the
question of the general solutions of the standard Boltzmann equation can be
tackled along two distinct lines. One is to attempt to solve Boltzmann
equation for the distribution $f_{1}$ itself in specific problems. Other is to
obtain new phenomenological equations in an approach initiated by Maxwell
\cite{maxwell} and continued by Grad [\cite{grad}, \cite{grad2}] (it was
called Grad's moments procedure \cite{grad1}). These moments produce
quantities with a clear physical meaning, namely, the densities of particles
and of energy and the fluxes of particles of first and second order in a
restricted fourteen-moments approach.

In brief, the $r$th-order moment is the flux of order $r$%
\begin{equation}
\mathbf{I}_{s}^{[r]}(\mathbf{r},t)=\int d^{3}p\text{ }u_{s}^{[r]}%
(\mathbf{p})f_{1}\left(  \mathbf{r},\mathbf{p};t\right)
\end{equation}
where $u_{s}^{[r]}$ is the $r$-rank tensor, $s\equiv n$ for particle motion
and $s\equiv h$ for energy motion,%
\begin{equation}
u_{n}^{[r]}(\mathbf{p})=\left[  \frac{\mathbf{p}}{m}...(r-times)...\frac
{\mathbf{p}}{m}\right]  \text{,}%
\end{equation}%
\[
u_{h}^{[r]}(\mathbf{p})=\frac{p^{2}}{2m}u_{n}^{[r]}(\mathbf{p})\text{,}%
\]
that is, the tensorial product of $r$-times the vector $\mathbf{p/}m$; for
$s\equiv n$, $r=0$ stands for the density, $r=1$ for the vector flux (or
current), $r=2$ for the flux of the flux which is related to the pressure
tensor field, and $r>2$ for all the other higher-order fluxes. For $s\equiv
h$, $r=0$ stands for the density of energy and $r\geq1$ for the respective
fluxes. The density of energy $h(\mathbf{r},t)$ follows from the trace of
$m\mathbf{I}_{n}^{[2]}$, namely
\begin{equation}
h(\mathbf{r},t)=\int d^{3}p\text{ }\frac{p^{2}}{2m}f_{1}\left(  \mathbf{r}%
,\mathbf{p};t\right)  \text{.}%
\end{equation}

The set composed by $n(\mathbf{r},t)$, $\mathbf{I}_{n}(\mathbf{r},t)$,
$\mathbf{I}_{n}^{[2]}(\mathbf{r},t)$ and $h(\mathbf{r},t)$ is the one
corresponding to Grad's fourteen moments approach. Finally, the hydrodynamic
equations are%
\begin{equation}
\frac{\partial}{\partial t}\mathbf{I}_{s}^{[r]}(\mathbf{r},t)=\int
d^{3}p\text{ }u_{s}^{[r]}(\mathbf{p})\frac{\partial}{\partial t}f_{1}\left(
\mathbf{r},\mathbf{p};t\right)  \label{eq131}%
\end{equation}
with $r=0,1,2,...,$ and where is to be introduced Eq.(\ref{eq24}). Equations
(\ref{eq131}) consists of an enormous set of coupled nonlinear
integro-differential equations. Evidently, it can be handled only in a
contracted version, introducing the hydrodynamics of order $0$, $1$, $2$,
etc..., thus classified according to the last flux that is retained in the
contraction of the description. Criteria for deciding the order of the
contraction must be established (see Ref. \cite{ramos1}). Hydrodynamics of
order zero leads for the density to satisfy Fick' standard diffusion equation,
the one of order one to Maxwell-Cattaneo equation, and the other orders to
generalized Burnett and super-Burnett equations.

To proceed further, and give a clear illustration of the functioning of the
theory, we consider the case of a solution of $N$ particles of mass $m$ (the
solute) in a fluid (the solvent) of $N_{R}$ particles of mass $M$. The former
is subjected to external forces -- driving it out of equilibrium --, and the
latter (the thermal bath) is taken in a steady state of constant equilibrium
with an external reservoir at temperature $T_{o}$. An analogous case, but at
the quantum mechanical level, is the one of carriers embedded in the ionic
lattice in doped or photoinjected semiconductors (see for example
\cite{luzzi3}-\cite{vasconcellos2}).

We write for the Hamiltonian%

\begin{equation}
\hat{H}=\hat{H}_{S}+\hat{H}_{R}+\hat{W}+\hat{H}_{P}\text{ ,} \label{eq10}%
\end{equation}
where, the first term on the right,
\begin{equation}
\hat{H}_{S}=\sum_{j=1}^{N}\frac{p_{j}^{2}}{2m}+\frac{1}{2}\sum_{j\neq k}%
^{N}V\left(  \left\vert \mathbf{r}_{j}-\mathbf{r}_{k}\right\vert \right)
\label{eq11}%
\end{equation}
is the Hamiltonian of the particles in the solute, consisting of their kinetic
energy and their pair interaction via a central force potential; the second
term is%

\begin{equation}
\hat{H}_{R}=\sum_{\mu=1}^{N_{R}}\frac{P_{\mu}^{2}}{2M}+\frac{1}{2}\sum
_{\mu\neq\nu=1}^{N_{R}}\Phi_{R}\left(  \left\vert \mathbf{R}_{\mu}%
-\mathbf{R}_{\nu}\right\vert \right)  \label{eq12}%
\end{equation}
which is the Hamiltonian of the particles in the solvent, acting as a thermal
bath, consisting of their kinetic energy plus their pair interaction via a
central force potential; moreover%

\begin{equation}
\hat{W}=\sum_{j=1}^{N}\sum_{\mu=1}^{N_{R}}w\left(  \left\vert \mathbf{r}%
_{j}-\mathbf{R}_{\mu}\right\vert \right)  \label{eq13}%
\end{equation}
is the interaction Hamiltonian of the particles with the thermal bath, and
$H_{P}=%
{\textstyle\sum_{i}}
V_{ext}(\mathbf{r}_{i},\mathbf{p}_{i},t)$ is the Hamiltonian associated to the
external force acting on the particles of the system.

Under the stated condition that the bath is in constant thermal equilibrium
with an external reservoir at temperature $T_{o}$, its macroscopic state is
characterized by the canonical distribution%

\begin{equation}
\varrho_{R}=Z^{-1}\exp\left\{  -\beta_{o}\hat{H}_{R}\right\}  \label{eq14}%
\end{equation}
where $\beta_{o}=\left[  k_{B}T_{o}\right]  ^{-1}$ and $Z$ is the
corresponding partition function. The auxiliary nonequilibrium statistical
operator of the whole system is the one of Eq. (\ref{eq3}) and Eq.
(\ref{eq4}). But, for simplicity, considering a dilute solution (large
distance in average between the particles) or that the potential $V$ is
screened (e.g., molecules in an ionized saline solvent, e.g. \cite{bouchaud}),
we can disregard the influence of the two particle potential, and then ignore
$\widehat{n}_{2}$, that is, taking $F_{2}=0$ in Eq. (\ref{eq4}) retaining only
$\widehat{n}_{1}$. In that case, we choose the single-particle reduced
density, $\widehat{n}_{1}(\mathbf{r},\mathbf{p}\mid\Gamma)$, as the only
relevant dynamical variable required. Hence, $\overline{\varrho}\left(
t,0\right)  $, of Eq. (\ref{eq4}), the auxiliary nonequilibrium statistical
operator for the particles embedded in the bath, is%
\[
\overline{\varrho}\left(  t,0\right)  =\exp\Bigg\{-\phi\left(  t\right)  -\int
d^{3}rd^{3}p\text{ }F_{1}\left(  \mathbf{r},\mathbf{p};t\right)
\]%
\[
\times\widehat{n}_{1}\left(  \mathbf{r},\mathbf{p}\right)  \Bigg\}
\]%
\begin{equation}
=%
{\displaystyle\prod\limits_{j=1}^{N}}
\overline{\varrho}_{j}\left(  t,0\right)  ,
\end{equation}
where%

\[
\overline{\varrho}_{j}\left(  t,0\right)  =\exp\left\{  -\phi_{j}\left(
t\right)  -\int d^{3}rd^{3}p\text{ }F_{1}\left(  \mathbf{r},\mathbf{p}%
;t\right)  \right.
\]%
\begin{equation}
\left.  \times\delta\left(  \mathbf{r}-\mathbf{r}_{j}\right)  \delta\left(
\mathbf{p}-\mathbf{p}_{j}\right)  \right\}  \label{eq16}%
\end{equation}
is a probability distribution for an individual particle, with $\phi\left(
t\right)  $ and $\phi_{j}\left(  t\right)  $ ensuring the normalization
conditions of $\overline{\varrho}$ and $\overline{\varrho}_{j}$.

The nonequilibrium equation of state \cite{luzzi2}, that is the one relating
the variables $f_{1}\left(  \mathbf{r},\mathbf{p};t\right)  $ and
$F_{1}\left(  \mathbf{r},\mathbf{p};t\right)  $ is%
\[
f_{1}\left(  \mathbf{r},\mathbf{p};t\right)  =Tr\left\{  \widehat{n}%
_{1}\left(  \mathbf{r},\mathbf{p}\right)  \overline{\varrho}\left(
t,0\right)  \right\}
\]%
\begin{equation}
=\exp\left\{  -F_{1}\left(  \mathbf{r},\mathbf{p};t\right)  \right\}  \text{,}
\label{eq17}%
\end{equation}
or%
\begin{equation}
F_{1}\left(  \mathbf{r},\mathbf{p};t\right)  =-\ln f_{1}\left(  \mathbf{r}%
,\mathbf{p};t\right)  \text{.} \label{eq18}%
\end{equation}
On the other hand, the evolution equation for $f_{1}$ following from the
NESEF-based kinetic theory, derived as shown in Ref. \cite{silva}, is the
generalized kinetic equation
\[
\frac{\partial}{\partial t}f_{1}\left(  \mathbf{r},\mathbf{p};t\right)
+\frac{\mathbf{P}\left(  \mathbf{r},\mathbf{p};t\right)  }{m}\cdot\nabla
f_{1}\left(  \mathbf{r},\mathbf{p};t\right)
\]%
\[
+\mathbf{F}\left(  \mathbf{r},\mathbf{p};t\right)  \cdot\nabla_{\mathbf{p}%
}f_{1}\left(  \mathbf{r},\mathbf{p};t\right)  -B\left(  \mathbf{p}\right)
f_{1}\left(  \mathbf{r},\mathbf{p};t\right)
\]%
\[
-A_{2}^{\left[  2\right]  }\left(  \mathbf{p}\right)  \odot\left[
\nabla_{\mathbf{p}}\nabla\right]  f_{1}\left(  \mathbf{r},\mathbf{p};t\right)
\]%
\begin{equation}
-B_{2}^{\left[  2\right]  }\left(  \mathbf{p}\right)  \odot\left[
\nabla_{\mathbf{p}}\nabla_{\mathbf{p}}\right]  f_{1}\left(  \mathbf{r}%
,\mathbf{p};t\right)  =0, \label{eq24}%
\end{equation}
obtained in the Markovian approximation \cite{luzzi}, \cite{zubarev2},
\cite{kuzemski1}, \cite{lauck}, where%
\begin{equation}
\frac{\mathbf{P}\left(  \mathbf{r},\mathbf{p};t\right)  }{m}=\frac{\mathbf{p}%
}{m}-\mathbf{A}_{1}\left(  \mathbf{p}\right)  \text{,} \label{eq25}%
\end{equation}%
\[
\mathbf{F}\left(  \mathbf{r},\mathbf{p};t\right)  =-\nabla V_{ext}\left(
\mathbf{r},\mathbf{p};t\right)  -\mathbf{B}_{1}\left(  \mathbf{p}\right)
\]%
\begin{equation}
-\mathbf{F}_{NL}\left(  \mathbf{r};t\right)  \text{,} \label{eq26}%
\end{equation}
with the explicit expressions, for the vectorial quantities $\mathbf{A}%
_{1}\left(  \mathbf{p}\right)  $, $\mathbf{B}_{1}\left(  \mathbf{p}\right)  $,
$\mathbf{F}_{NL}\left(  \mathbf{r};t\right)  $, the second-rank
tensors\ $A_{2}^{\left[  2\right]  }\left(  \mathbf{p}\right)  $,
$B_{2}^{\left[  2\right]  }\left(  \mathbf{p}\right)  $, and the
scalar\ $B\left(  \mathbf{p}\right)  $, together with a description of the
physical meaning of the several contributions, are given in Ref. \cite{silva}.
We also wrote the symbol $\odot$ for full contraction of tensors.

The distribution $f_{1}\left(  \mathbf{r},\mathbf{p};t\right)  $ that follows
solving Eq.(\ref{eq24}) provides a complete information about the actual
distribution of particles, and therefore of the physical properties of the
system. Alternatively, if one knows all the moments of the distribution we do
have a knowledge of its characteristics. A knowledge of some moments is not
sufficient to determine the distribution completely; it implies in only
possessing partial knowledge on the characteristics of this distribution
\cite{reif} (this is related to Tchebychef's procedure for obtaining
characteristics of a probability distribution when we do possess the moments
of successive order, e.g. \cite{castelnuovo}). On this H. Grad noticed that
the question of the general solutions of the standard Boltzmann kinetic
equation can be tackled along two distinct lines. One is to attempt to solve
Boltzmann equation for the distribution $f_{1}$ itself in specific problems.
Other is to obtain new phenomenological equations which generalize the usual
(classical-Onsagerian) fluid dynamical equations. The object is to show the
transition from the Boltzmann equation in which a state is given by
$f_{1}\left(  \mathbf{r},\mathbf{p};t\right)  $ to the conventional fluid
description in which a state is given by the density $n\left(  \mathbf{r}%
,t\right)  $, the velocity field $\mathbf{v}\left(  \mathbf{r},t\right)  $ ,
and the stress tensor $T^{\left[  2\right]  }\left(  \mathbf{r},t\right)  $,
in a sufficient generality to cover a broad class of problems. This approach
was initiated by Maxwell \cite{maxwell} and continued by Grad \cite{grad} (it
was called Grad's moments procedure) \cite{grad1}.

We do perform here an extensive generalization of the moments procedure,
consisting into introducing the full set of moments of $f_{1}\left(
\mathbf{r},\mathbf{p};t\right)  $, of Eq.(\ref{eq24}), in the variable
$\mathbf{p}$. These moments produce quantities with a clear physical meaning,
namely, the densities of particles and of energy and their fluxes of all
order: the two vectorial fluxes, or currents, the tensorial fluxes, beginning
with the second-order one which is the flux of the first-order flux (the
current of particles) which is related to the pressure tensor, and all the
other higher-order fluxes. In that way we obtain a quite generalized
\textit{Mesoscopic Hydrodynamics coupled to a Non-Equilibrium Thermodynamics},
all together in the kinetic approach provided by NESEF, as described in next section.

\section{Mesoscopic Hydro-Thermodynamics in NESEF.}

Let us introduce, in the variable $\mathbf{p}$, the moments of the
distribution $f_{1}\left(  \mathbf{r},\mathbf{p};t\right)  $, namely
\begin{equation}
n(\mathbf{r},t)=\int d^{3}p\text{ }f_{1}\left(  \mathbf{r},\mathbf{p}%
;t\right)  \text{,} \label{eq28}%
\end{equation}
which is the density of particles;%
\begin{equation}
\mathbf{I}_{n}(\mathbf{r},t)=\int d^{3}p\text{ }\mathbf{u}\left(
\mathbf{p}\right)  f_{1}\left(  \mathbf{r},\mathbf{p};t\right)  \text{,}
\label{eq29}%
\end{equation}
where
\begin{equation}
\mathbf{u}\left(  \mathbf{p}\right)  =\mathbf{p/}m\text{,} \label{eq30}%
\end{equation}
with $\mathbf{I}_{n}$ being the flux (current) of particles;
\begin{equation}
I_{n}^{[2]}(\mathbf{r},t)=\int d^{3}p\text{ }\mathbf{u}^{[2]}\left(
\mathbf{p}\right)  f_{1}\left(  \mathbf{r},\mathbf{p};t\right)  \label{eq31}%
\end{equation}
where $\mathbf{u}^{[2]}=[\mathbf{uu}]$ is the tensorial product of vectors
$\mathbf{u}$, with $I_{n}^{[2]}$ being the second-order flux \ (or flux of the
first flux), a rank-2 tensor, which multiplied by the mass is related to the
pressure tensor and%
\begin{equation}
I_{n}^{[l]}(\mathbf{r},t)=\int d^{3}p\text{ }\mathbf{u}^{[l]}\left(
\mathbf{p}\right)  f_{1}\left(  \mathbf{r},\mathbf{p};t\right)  \label{eq32}%
\end{equation}
are the higher-order fluxes of order $l\geq3$ (the previous three of Eqs.
(\ref{eq28}), (\ref{eq29}) and (\ref{eq31}) are those for $l=0$, $1$ and $2$
respectively), where $u^{[l]}$ is the $l$-rank tensor consisting of the
tensorial product of $l$ vectors $\mathbf{u}$ of Eq. (\ref{eq30}) that is,%
\begin{equation}
\mathbf{u}^{[l]}(\mathbf{p})=\left[  \frac{\mathbf{p}}{m}\frac{\mathbf{p}}%
{m}...(l-times)...\frac{\mathbf{p}}{m}\right]  . \label{eq33}%
\end{equation}
\ \ We do have what can be called the \textit{family of hydrodynamical
variables describing the material motion}, i. e., the set%
\begin{equation}
\left\{  n(\mathbf{r},t),\text{ }\mathbf{I}_{n}(\mathbf{r},t),\text{ }%
\{I_{n}^{[l]}(\mathbf{r},t)\}\right\}  \label{eq34}%
\end{equation}
with $l=2,$ $3,....$

On the other hand, we have the \textit{family of hydrodynamical variables
describing the thermal motion}, consisting of
\begin{equation}
h(\mathbf{r},t)=\int d^{3}p\text{ }\frac{p^{2}}{2m}f_{1}\left(  \mathbf{r}%
,\mathbf{p};t\right)  , \label{eq35}%
\end{equation}%
\begin{equation}
\mathbf{I}_{h}(\mathbf{r},t)=\int d^{3}p\frac{p^{2}}{2m}\text{ }%
\frac{\mathbf{p}}{m}f_{1}\left(  \mathbf{r},\mathbf{p};t\right)  ,
\label{eq36}%
\end{equation}%
\begin{equation}
I_{h}^{[l]}(\mathbf{r},t)=\int d^{3}p\frac{p^{2}}{2m}\text{ }\mathbf{u}%
^{[l]}\left(  \mathbf{p}\right)  f_{1}\left(  \mathbf{r},\mathbf{p};t\right)
, \label{eq37}%
\end{equation}
with $l=2$, $3$, $...$; that is, in compact form those in the set
\begin{equation}
\left\{  h(\mathbf{r},t),\text{ }\mathbf{I}_{h}(\mathbf{r},t),\text{ }%
\{I_{h}^{[l]}(\mathbf{r},t)\}\right\}  , \label{eq38}%
\end{equation}
which are, respectively, the density of energy, its first vectorial flux
(current), and the higher-order tensorial fluxes. It can be noticed that in
this case of a parabolic type energy-momentum dispersion relation,
$E(p)=p^{2}/2m,$ the set of Eq. (\ref{eq38}) is encompassed in the previous
one: In fact%
\begin{equation}
h(\mathbf{r},t)=\frac{m}{2}Tr\left\{  I_{n}^{[2]}(\mathbf{r},t)\right\}
\label{eq39}%
\end{equation}%
\begin{equation}
\mathbf{I}_{h}(\mathbf{r},t)=\frac{m}{2}Tr_{2}\left\{  I_{n}^{[3]}%
(\mathbf{r},t)\right\}  \text{ ,} \label{eq40}%
\end{equation}
where $Tr_{2}$ stands for the contraction of two indexes, and, in general
\begin{equation}
I_{h}^{\left[  l\right]  }(\mathbf{r},t)=\frac{m}{2}Tr_{2}\left\{
I_{n}^{[l+2]}(\mathbf{r},t)\right\}  \text{,} \label{eq41}%
\end{equation}
for all the other higher-order fluxes of energy. Hence, any flux of energy of
order $l$ is contained in the flux of matter of order $l+2$. In what follows
we concentrate the attention on the study of the hydrodynamic motion of the
particles, with heat transport to be dealt with in a future communication in
this series.

\section{MHT Evolution Equations in NESEF}

We proceed now to the derivation of the MHT equations, that is, the equations
of evolutions for the basic macrovariables of the family of material motion,
i. e. those in set of Eq.(\ref{eq34}).

Let us consider the flux of order $l$ ($l=0,1,2,...$); its evolution equation
is
\begin{equation}
\frac{\partial}{\partial t}I_{n}^{[l]}(\mathbf{r},t)=\int d^{3}p\text{
}u^{[l]}\left(  \mathbf{p}\right)  \frac{\partial}{\partial t}f_{1}\left(
\mathbf{r},\mathbf{p};t\right)  . \label{eq42}%
\end{equation}
Using Eq. (\ref{eq24}), but excluding a dependence on $\mathbf{p}$ of the
external force, after lengthy but straightforward calculations we arrive to
the general set of coupled equations for the density, $l=0$, the current,
$l=1$, and all the other higher-order fluxes, $l\geq2$, given by%

\[
\frac{\partial}{\partial t}I_{n}^{[l]}(\mathbf{r},t)+\nabla\cdot I_{n}%
^{[l+1]}(\mathbf{r},t)=
\]%
\[
=\frac{1}{m}%
{\displaystyle\sum\limits_{s=1}^{l}}
{\huge \wp}(1,s)\left[  \boldsymbol{%
\mathcal{F}%
}_{ext}\left(  \mathbf{r},t\right)  I_{n}^{[l-1]}(\mathbf{r},t)\right]
\]%
\begin{equation}
+J_{\tau}^{[l]}(\mathbf{r},t)+J_{L}^{[l]}(\mathbf{r},t)+J_{NL}^{[l]}%
(\mathbf{r},t), \label{eq43}%
\end{equation}
where ${\huge \wp}(1,s)$ means that we must take a permutation of the first
free index (1) with the s-th (s=1,2,3,...,$l$) free index of the Cartesian
tensor $\left[  \boldsymbol{%
\mathcal{F}%
}_{ext}\left(  \mathbf{r},t\right)  I_{n}^{[l-1]}(\mathbf{r},t)\right]  $,
when written in the indicial notation. Observe that the number of terms in the
sum is given by the number of all permutations of $l$ symbols in which $l-1$
are repeated. All this ensures the correct symmetry of this contribution, that
is, a fully symmetrical tensor of order $l$.

The several terms on the right of Eq. (\ref{eq43}) are,%
\begin{equation}
\boldsymbol{%
\mathcal{F}%
}_{ext}\left(  \mathbf{r},t\right)  =-\nabla V_{ext}\left(  \mathbf{r}%
;t\right)  , \label{eq44}%
\end{equation}
that is the applied external force, created by the action of the potential
$V_{ext}$, and the terms $\mathbf{A}_{1}\left(  \mathbf{p}\right)  $,
$\mathbf{B}_{1}\left(  \mathbf{p}\right)  $ and $\mathbf{F}_{NL}\left(
\mathbf{r};t\right)  $ present in Eqs. (\ref{eq25}) and (\ref{eq26}) have
respectively been incorporated into $J_{\tau}^{[l]}$, $J_{L}^{[l]}$ and
$J_{NL}^{[l]}$, which are,%

\[
J_{\tau}^{[l]}(\mathbf{r},t)=\frac{n_{R}}{\mathcal{V}}\frac{\sqrt{2\pi
M\beta_{o}}}{m^{2}}\sum_{\mathbf{Q}}\frac{\left\vert \psi\left(  Q\right)
\right\vert ^{2}}{Q}\times
\]%
\[
\int d^{3}p%
{\displaystyle\sum\limits_{P}}
\left[  \mathbf{QQu}^{[l-2]}(\mathbf{p})\right]
\]%
\[
\times\exp\left[  {\normalsize -\alpha}\left(  \frac{\mathbf{Q}}{Q}%
\cdot\mathbf{p}\right)  ^{2}\right]  f_{1}\left(  \mathbf{r},\mathbf{p}%
;t\right)
\]%
\begin{equation}
-\frac{n_{R}}{\mathcal{V}}\frac{\left(  M\beta_{o}\right)  ^{3/2}\pi}%
{\sqrt{2\pi}m^{2}}\sum_{\mathbf{Q}}\frac{\left\vert \psi\left(  Q\right)
\right\vert ^{2}}{Q}\left(  \frac{1}{M}+\frac{1}{m}\right)  \times\nonumber
\end{equation}%
\[
\int d^{3}p{}%
{\displaystyle\sum\limits_{P}}
\left[  \mathbf{Qu}^{[l-1]}(\mathbf{p})\right]  \mathbf{Q}\cdot\mathbf{p}%
\]%
\begin{equation}
\times\exp\left[  -\alpha\left(  \frac{\mathbf{Q}}{Q}\cdot\mathbf{p}\right)
^{2}\right]  f_{1}\left(  \mathbf{r},\mathbf{p};t\right)  , \label{eq46}%
\end{equation}
with $\alpha=M\beta_{o}/2m^{2}$, $\psi\left(  Q\right)  $ is the Fourier
transform of the interaction potential between the particles and the thermal
bath, i. e., Eq.(\ref{eq13}) and according to Eq. (\ref{eq33})
\begin{equation}
\left[  \mathbf{QQu}^{[l-2]}(\mathbf{p})\right]  =\left[  \mathbf{QQ}%
\frac{\mathbf{p}}{m}...\left(  l-2\right)  \text{ }times\text{ }%
...\frac{\mathbf{p}}{m}\right]  ,
\end{equation}%
\begin{equation}
J_{NL}^{[l]}\left(  \mathbf{r},t\right)  =-\frac{1}{m}%
{\displaystyle\sum\limits_{s=1}^{l}}
{\huge \wp}(1,s)\left[  \mathbf{F}_{NL}\left(  \mathbf{r},t\right)
I_{n}^{[l-1]}(\mathbf{r},t)\right]  , \label{eq48}%
\end{equation}%
\[
\mathbf{F}_{NL}\left(  \mathbf{r};t\right)  =\int d^{3}r^{\prime}\int
d^{3}p^{\prime}\mathbf{G}_{NL}\left(  \mathbf{r}^{\prime}-\mathbf{r}%
,\mathbf{p}^{\prime}\right)
\]%
\begin{equation}
f_{1}\left(  \mathbf{r}^{\prime},\mathbf{p}^{\prime};t\right)  , \label{eq481}%
\end{equation}%
\[
\mathbf{G}_{NL}\left(  \mathbf{r}^{\prime}-\mathbf{r},\mathbf{p}^{\prime
}\right)  =\frac{n_{R}\beta_{o}}{\mathcal{V}}%
{\textstyle\sum\limits_{\mathbf{Q}}}
\mathbf{Q}\left\vert \psi\left(  Q\right)  \right\vert ^{2}%
\]%
\[
\times\Big\{iF\left(  \mathbf{Q},\mathbf{p}^{\prime}\right)
\]%
\[
+\left(  \frac{M\beta_{o}}{2\pi}\right)  ^{1/2}\frac{\pi}{m}\frac
{\mathbf{Q}\cdot\mathbf{p}^{\prime}}{Q}\times\exp\left[  -\alpha\left(
\frac{\mathbf{Q}}{Q}\cdot\mathbf{p}^{\prime}\right)  ^{2}\right]  \Big\}
\]%
\begin{equation}
\times\exp\left[  i\mathbf{Q\cdot(r}^{\prime}-\mathbf{r)}\right]  ,
\label{eq482}%
\end{equation}

\[
J_{L}^{[l]}(\mathbf{r},t)=-\frac{n_{R}}{\mathcal{V}}\frac{M\beta_{o}}{m^{2}%
}\sum_{\mathbf{Q}}\frac{\left\vert \psi\left(  Q\right)  \right\vert ^{2}%
}{Q^{2}}%
\]%
\[
\times\mathbf{Q}\cdot\nabla\int d^{3}p\sum_{P}\left[  \mathbf{Qu}%
^{[l-1]}(\mathbf{p})\right]  F(\mathbf{p,Q})
\]%
\begin{equation}
\times f_{1}\left(  \mathbf{r},\mathbf{p};t\right)  \text{,} \label{eq49}%
\end{equation}
where%

\begin{equation}
F(\mathbf{p,Q})=1+%
{\textstyle\sum\limits_{n=1}^{\infty}}
\frac{(-1)^{n}}{\left(  2n-1\right)  !!}2^{n}\alpha^{n}\left(  \frac
{\mathbf{Q}}{Q}\cdot\mathbf{p}\right)  ^{2n}. \label{eq47}%
\end{equation}
Observe that the notation%

\[
\sum_{P}\left[  \mathbf{Qu}^{[l-1]}(\mathbf{p})\right]
\]%
\begin{equation}
\equiv\sum_{P}\left[  \mathbf{Q}\frac{\mathbf{p}}{m}...\left(  l-1\right)
\text{ }times\text{ }...\frac{\mathbf{p}}{m}\right]  \label{eq471}%
\end{equation}
means that one has to sum all permutations of the vectors in \ order to ensure
that the tensor has the same symmetry of the tensor $I_{n}^{[l]}$ on the left
hand side of Eq. (\ref{eq43}).

Next, making a Taylor series expansion of the exponential in both
contributions in Eq. (\ref{eq46}) and in Eq. (\ref{eq482}), i. e.,
\begin{equation}
\exp\left[  -\alpha\left(  \frac{\mathbf{Q}}{Q}\cdot\mathbf{p}\right)
^{2}\right]  =\sum_{k=0}^{\infty}\frac{\left(  -1\right)  ^{k}\alpha^{k}}%
{k!}\mathbf{Q}^{\left[  2k\right]  }\odot\mathbf{p}^{\left[  2k\right]
}\text{,}%
\end{equation}
where $\mathbf{Q}^{\left[  l\right]  }$ stands for
\begin{equation}
\mathbf{Q}^{\left[  l\right]  }=\left[  \mathbf{Q}...\left(  l-times\right)
...\mathbf{Q}\right]  ,
\end{equation}
and we recall that $\odot$ stands for full contraction of the two tensors of
rank $2k$. Then using Eq.(\ref{eq47}) for $F(\mathbf{p,Q})$, we can rewrite
Eqs. (\ref{eq46}), (\ref{eq48}) and (\ref{eq49}) in a closed form in terms of
all the fluxes, namely%

\[
{\large J}_{{\large \tau}}^{{\large [l]}}{\large (\mathbf{r},t)}{\large =}%
\sum_{{\large k=0}}^{{\large \infty}}\widehat{{\Large \wp}}_{l}\left[
{\large \Lambda}_{\tau o}^{{\large [2k+2]}}\odot{\large I}_{n}%
^{{\large [2k+l-2]}}{\large (\mathbf{r},t)}\right]
\]%
\begin{equation}
+\sum_{{\large k=0}}^{{\large \infty}}\sum_{{\large s=1}}^{{\large l}%
}{\huge \wp}(1,s)\left[  {\large \Lambda}_{\tau}^{{\large [2k+2]}}%
\odot{\large I}_{n}^{{\large [2k+l]}}{\large (\mathbf{r},t)}\right]  ,
\label{eq50}%
\end{equation}%
\[
{\large J}_{L}^{[l]}{\large (\mathbf{r},t)=}\sum_{{\large k=0}}%
^{{\large \infty}}\sum_{{\large s=1}}^{{\large l}}{\huge \wp}(1,s)\left[
{\large \Lambda}_{L}^{{\large [2k+2]}}\right.
\]

\begin{equation}
\left.  \odot{\large \nabla I}_{n}^{{\large [2k+l-1]}}{\large (\mathbf{r}%
,t)}\right]  \text{,} \label{eq52}%
\end{equation}%
\[
J_{NL}^{[l]}\left(  \mathbf{r},t\right)  {\large =}{\large -}\frac{1}{m}%
\sum_{s{\large =1}}^{{\large l}}{\huge \wp}(1,s)\sum_{{\large k=0}%
}^{{\large \infty}}\int d^{3}\mathbf{r}^{\prime}%
\]%
\[
\left\{  \left[  \left[  {\large \Lambda}_{NL1}^{{\large [2k+1]}}\left(
{\large \mathbf{r}}^{\prime}-{\large \mathbf{r}}\right)  \odot{\large I}%
_{n}^{{\large [2k]}}{\large (\mathbf{r}}^{\prime}{\large ,t)}\right]
I_{n}^{[l-1]}(\mathbf{r},t)\right]  \right.
\]%
\begin{equation}
+\left.  \left[  \left[  {\large \Lambda}_{NL2}^{{\large [2k+2]}}\left(
{\large \mathbf{r}}^{\prime}-{\large \mathbf{r}}\right)  \odot{\large I}%
_{n}^{{\large [2k+1]}}{\large (\mathbf{r}}^{\prime}{\large ,t)}\right]
I_{n}^{[l-1]}(\mathbf{r},t)\right]  \right\}  \text{, } \label{eq53}%
\end{equation}
where%

\[
\widehat{{\Huge \wp}}_{l}{\huge =}\sum_{{\large s=2}}^{{\large l}}{\huge \wp
}(2,s)+\sum_{r{\large =3}}^{{\large l}}{\huge \wp}(1,r)
\]%
\begin{equation}
+\sum_{r{\large =3}}^{{\large l-1}}\sum_{{\large s=r+1}}^{{\large l}%
}{\huge \wp}(1,r;2,s)
\end{equation}
is an operator involving the set of permutations that ensures the proper
symmetry of the tensor on which it acts. Here the operation ${\large \Lambda
}^{[r]}\odot{\large \Lambda}^{[s]}$ indicates the contraction of some indexes
in order to give the right tensorial order of the equation. For example in
Eqs. (\ref{eq50}) and (\ref{eq52}) it indicates the contraction of $(r+s-l)/2$
indexes as to produce a tensor of rank $l$. The several tensorial kinetic
coefficients are given in Appendix A: Eqs. (\ref{eqa1}) to Eq. (\ref{eqa8}).

We can see that the expressions for $J_{\tau}^{[l]}$ and $J_{L}^{[l]}$ are
linear in the hydrodynamic basic variables of the set of Eq. (\ref{eq34}) with
tensorial coefficients ${\large \Lambda}^{[r]}$. The first one contains
contributions of a relaxation character, and the second involves local
couplings with the different fluxes. On the other hand ${\large J}%
_{{\large NL}}^{{\large [l]}}$ is nonlinear (bilinear) in the fluxes
accounting for nonlocal correlations involving all of them. Next, we
reorganize these expressions setting into evidence \ the contributions that
contain the neighboring fluxes to the one of order $l$, namely $I_{n}%
^{[l-1]}(\mathbf{r},t)$ and $I_{n}^{[l+1]}(\mathbf{r},t)$, that is, arising
out of the terms with $k=0$ and $k=1$ in the sum over $k$ in Eqs. (\ref{eq50})
and (\ref{eq52}), to obtain Eq. (\ref{eq43}) in the form,%
\begin{equation}
\frac{\partial}{\partial t}{\large I}_{n}^{[l]}{\large (\mathbf{r}%
,t)+\nabla\cdot I}_{n}^{[l+1]}{\large (\mathbf{r},t)}{\large =}\nonumber
\end{equation}%
\[
=\frac{1}{m}\sum_{{\large s=1}}^{{\large l}}{\huge \wp}(1,s)\left[
\boldsymbol{%
\mathcal{F}%
}_{ext}\left(  \mathbf{r},t\right)  I_{n}^{[l-1]}(\mathbf{r},t)\right]
\]%
\begin{equation}
-{\large \theta}_{l}^{-1}{\large I}_{n}^{[l]}{\large (\mathbf{r}%
,t)}+{\Large a}_{{\Large Lo}}\sum_{{\large s=1}}^{{\large l}}{\huge \wp
}(1,s)\left[  \nabla I_{n}^{[l-1]}(\mathbf{r},t)\right] \nonumber
\end{equation}%
\[
+2l{\Large a}_{{\Large L1}}{\large \nabla\cdot I}_{n}^{[l+1]}%
{\large (\mathbf{r},t)+}{\Large b}_{{\Large \tau o}}\left\{
\widehat{{\Huge \wp}}_{l}\left[  \mathbf{1}^{\left[  2\right]  }I_{n}%
^{[l-2]}(\mathbf{r},t)\right]  \right\}
\]%
\begin{equation}
{\large +J_{NL}^{[l]}\left(  \mathbf{r},t\right)  +S}_{n}^{\left[  l\right]
}{\large (\mathbf{r},t)}\text{.} \label{eq59}%
\end{equation}

The last term on the right of Eq. (\ref{eq59}) is given by,%

\[
{\large S}_{n}^{\left[  l\right]  }{\large (}\mathbf{r}{\large ,t)}%
{\large =}{\Large b}_{{\Large \tau1}}\frac{2}{m}\left\{  \widehat{{\Huge \wp}%
}_{l}\left[  \mathbf{1}^{\left[  2\right]  }I_{h}^{[l-2]}(\mathbf{r}%
,t)\right]  \right\}
\]

\[
+3l{\Large a}_{{\Large \tau1}}\frac{2}{m}I_{h}^{[l]}(\mathbf{r},t)
\]%
\[
+{\Large a}_{{\Large L1}}\frac{2}{m}\sum_{{\large s=1}}^{{\large l}}%
{\huge \wp}{\large (}1{\large ,s)}\left[  \nabla I_{h}^{[l-1]}(\mathbf{r}%
,t)\right]
\]%
\begin{equation}
+{\large R}_{n}^{\left[  l\right]  }{\large (}\mathbf{r}{\large ,t)}\text{,}
\label{eq60}%
\end{equation}
where ${\large R}_{n}^{\left[  l\right]  }{\large (\mathbf{r},t)}$, given in
Appendix A, contains the contributions of the fluxes of order higher than
$l+2$; and the kinetic coefficients ${\Large a}_{{\Large \tau1}}$,
${\Large a}_{{\Large Lo}}$, ${\Large a}_{{\Large L1}}$, ${\Large b}%
_{{\Large \tau o}}$, ${\Large b}_{{\Large \tau1}}$ are given in Appendix B.
The first three contributions on the right of Eq. (\ref{eq60}) are associated
to the fluxes of energy of orders $l-2$, $l-1$ and $l$, terms that can be
considered consisting of thermo-striction effects and which, then, couple
these equations with the set of kinetic equations describing the movement of
energy, i. e. the hydrodynamical variables of the set of Eq. (\ref{eq38}).
However it can be noticed the already mentioned fact that the fluxes of energy
can be given in terms of those of particles, namely, $I_{h}^{\left[  l\right]
}=(m/2)Tr_{2}\left\{  I_{n}^{[l+2]}\right\}  $, [ cf. Eqs. (\ref{eq39}) to
(\ref{eq41})]. Moreover,
\begin{equation}
{\large \theta}_{l}^{-1}=l\left[  \left\vert {\Large a}_{{\Large \tau o}%
}\right\vert +\left(  l-1\right)  \left\vert {\Large b}_{{\Large \tau1}%
}\right\vert \right]  , \label{eq61}%
\end{equation}
with ${\large \theta}_{l}$ playing the role of a Maxwell-characteristic time
for the $l$-$th$ flux.

We stress that $l=0$ corresponds to the density $n(\mathbf{r},t)$, $l=1$ to
the first flux (current) $\mathbf{I}_{n}(\mathbf{r},t)$, $l=2$ to the second
flux $I_{n}^{[2]}(\mathbf{r},t)$ which multiplied by the mass $m$ is related
to the pressure tensor $P^{[2]}(\mathbf{r},t)$], $l>2$ to the other
higher-order fluxes. Hence, Eq. (\ref{eq59}) represents the coupled set of
evolution equations involving the density and all its fluxes in its most
general form. It must be noticed that it is linear in the basic variables; no
approximation has been introduced. Nonlinearities should arise out of the
interparticle interaction inclusion which we have disregarded in the present
communication (case of a dilute solution). However, as already noticed, such
set of equations is intractable, and, of course, we need to look in each case
on how to find the best description using the smallest possible numbers of
variables. In other words to introduce an appropriate -- for each case --
contraction of description: \textit{this contraction implies in retaining the
information considered as relevant for the problem in hands, and to disregard
nonrelevant information }\cite{balian}\textit{.}

Elsewhere \cite{ramos1} we have discussed the question of the contraction of
description (reduction of the dimensions of the nonequilibrium thermodynamic
space of states). As shown, a criterion for justifying the different levels of
contraction is derived: It depends on the range of wavelengths and frequencies
which are relevant for the characterization, in terms of normal modes, of the
hydro-thermodynamic motion in the nonequilibrium open system. Maxwell times
have a particular relevance then we proceed to analyze them.

\section{The Hierarchy of Maxwell Characteristic Times}

Let us now analyze Maxwell characteristics times of Eq. (\ref{eq61}) \ to show
that they follow a hierarchy of values. First we note that, taking into
account \ Eq. (\ref{eq117}) into Eq. (\ref{eq61}) and the definition in Eq.
(\ref{eqb1}) we can write%
\begin{equation}
\theta_{l}^{-1}=l\left[  1+\frac{1}{5}\frac{M}{m+M}\left(  l-1\right)
\right]  \theta_{1}^{-1} \label{eq62}%
\end{equation}
what tell us that any characteristic time for $l\geq2$ is proportional to the
one of $l=1$, that is the one for the first flux which multiplied by the mass
$m$ is the linear momentum density and then all \ are proportional to the
linear momentum relaxation time. On the other hand we do have that%

\begin{equation}
\frac{\theta_{l+1}}{\theta_{l}}=\frac{l}{l+1}\frac{5\left(  1+x\right)
+l-1}{5\left(  1+x\right)  +l}\text{,} \label{eq63}%
\end{equation}
for $l=1,2,3,...$ and where $x=m/M$, and then the ordering sequence%

\begin{equation}
\theta_{1}>\theta_{2}>\theta_{3}>...>\theta_{l}>\theta_{l+1}>...\text{.}
\label{eq64}%
\end{equation}
is verified which can be considered to represent a kind of Bogoliubov%
\'{}%
s hierarchy of characteristic times \cite{bogoliubov} in generalized
hydrodynamics, and we can see that $\theta_{l}\rightarrow0$ as $l\rightarrow
\infty$. Moreover according to Eq. (\ref{eq62}) it follows that%
\begin{equation}
\theta_{l}=\frac{5\left(  1+x\right)  }{l\left[  5\left(  1+x\right)
+(l-1)\right]  }\theta_{1}\text{.} \label{eq65}%
\end{equation}

Comparing with the second flux $\left(  l=2\right)  $, the one related to the
pressure tensor, it follows that for the Brownian particles $\left(
x>>1\right)  $ $\theta_{2}\simeq\theta_{1}/2$ and for the Lorentz particles
$\left(  x<<1\right)  $ $\theta_{2}\simeq5\theta_{1}/12$. A comparison with
the third flux leads to the results that $\theta_{3}\simeq\theta_{1}/3$ and
$\theta_{3}\simeq5\theta_{1}/21$ for the Brownian and Lorentz particles
respectively. For any $l$ we do have approximately:

1) for the Brownian particle $\left(  m/M>>1\right)  $
\begin{equation}
\theta_{l}\simeq\theta_{1}/l\text{,}%
\end{equation}
\ \ 2) for the Lorentz particle $\left(  m/M<<1\right)  $
\begin{equation}
\theta_{l}\simeq\left[  5/\left(  4+l\right)  l\right]  \theta_{1}\text{,}%
\end{equation}
\ or $\theta_{l}\simeq5\theta_{1}/l^{2}$ for large $l$. \ 

Moreover according to Eq. (\ref{eq63}) as the order of flux largely increases
its characteristic Maxwell time approaches zero, and $\theta_{l+1}/\theta
_{l}\simeq1$, with both being practically null. In Fig. 1 it is displayed the
ratio of characteristic Maxwell times, for flux of order $\ell$ with the
momentum relaxation time, as a function of $m/M$, i. e., the quotient of the
masses $m$ of the system and $M$ of the thermal bath.%

{\includegraphics[
natheight=3.199800in,
natwidth=4.199500in,
height=2.1024in,
width=2.751in
]%
{LUF10D00.wmf}%
}

Figure 1: The quotient between several Maxwell characteristic times and the
one of the first flux as a function of $m/M.$

Observe in figure--1 that the quotient of masses has little effect on
$\theta_{l}/\theta_{1}$ but it varies significantly with the order $\ell$ of
the flux and plays a particular relevance in the criteria for stablishing the
contraction of description discussed below.

\section{ On the Criterion of Contracted Description and an Application}

Returning to the question of the contracted description it can be shown
\cite{ramos1} that a truncation criterion can be derived, which \textit{rests
on the characteristics of the hydrodynamic motion that develops under the
given experimental procedure.}

Since inclusion of higher and higher order fluxes implies in describing a
motion involving increasing Knudsen numbers per hydrodynamic mode (that is
governed by smaller and smaller wavelengths -- larger and larger wavenumbers
-- accompanied by higher and higher frequencies), in a qualitative manner, we
can say that, as a general \ \textquotedblleft thumb rule,\textquotedblright%
\ the criterion indicates that \textit{a more and more restricted contraction
can be used when larger and larger are the prevalent wavelengths in the
motion. }Therefore, in simpler words, when the motion becomes more and more
smooth in space and time, the more reduced can be the dimension of the basic
macrovariables space to be used for the description of the nonequilibrium
thermodynamic state of the system.

As shown elsewhere \cite{ramos1}, it can be conjectured \ a general
contraction criterion, namely, a contraction of order \textit{r} (meaning
keeping the densities and their fluxes \ up to order \textit{r}) can be
introduced, once we can show that in the spectrum of wavelengths, which
characterize the motion, predominate those larger than a \textquotedblleft
frontier\textquotedblright\ one, $\lambda_{\left(  r,r+1\right)  }^{2}%
=v^{2}\theta_{r}\theta_{r+1}$, where $v$ is of the order of the thermal
velocity and $\theta_{r}$ and $\theta_{r+1}$ the corresponding Maxwell times.
We shall try to illustrate the matter using a contraction of order $2$.

Let us first write down the equations of evolution, whose general expression
is \ given in Eq. (\ref{eq59}), corresponding to the density and its fluxes of
all order, for $\ell=0$: the density, for $\ell=1$: the first flux of the
density, $\ell=2$: the flux of the first flux which multiplied by $m$ is the
pressure tensor field, and $\ell=3$: the flux of the pressure. We do have,
respectively,%
\begin{equation}
\frac{\partial}{\partial t}n(\mathbf{r},t)+\nabla\cdot\mathbf{I}%
_{n}(\mathbf{r},t)=0, \label{eq68}%
\end{equation}

\[
\frac{\partial}{\partial t}\mathbf{I}_{n}{\large (\mathbf{r},t)+\nabla\cdot
I}_{n}^{[2]}{\large (\mathbf{r},t)}{\large =}\frac{n(\mathbf{r},t)}%
{m}\boldsymbol{%
\mathcal{F}%
}\left(  \mathbf{r},t\right)
\]%
\[
-\theta_{1}^{-1}\mathbf{I}_{n}{\large (\mathbf{r},t)}+{\Large a}_{{\Large Lo}%
}\nabla n(\mathbf{r},t)
\]%
\begin{equation}
+2{\Large a}_{L1}\nabla\cdot I_{n}^{[2]}(\mathbf{r},t)+\mathbf{S}_{n}\left(
\mathbf{r},t\right)  , \label{eq69}%
\end{equation}

\[
\frac{\partial}{\partial t}{\Large I}_{n}^{\left[  2\right]  }%
{\large (\mathbf{r},t)+\nabla\cdot I}_{n}^{[3]}{\large (\mathbf{r},t)}%
\]%
\[
=\frac{1}{m}\left\{  \left[  \boldsymbol{%
\mathcal{F}%
}\left(  \mathbf{r},t\right)  \mathbf{I}_{n}{\large (\mathbf{r},t)]+[}\text{
}\mathbf{I}_{n}{\large (\mathbf{r},t)}\text{ }\boldsymbol{%
\mathcal{F}%
}\left(  \mathbf{r},t\right)  \right]  \right\}
\]%
\[
-\theta_{2}^{-1}{\large I}_{n}^{[2]}{\large (\mathbf{r},t)+}{\Large a}%
_{Lo}\left\{  \nabla\mathbf{I}_{n}(\mathbf{r},t)+\left[  \nabla\mathbf{I}%
_{n}(\mathbf{r},t)\right]  ^{tr}\right\}
\]%
\begin{equation}
+4\text{ }{\Large a}_{L1}\nabla\cdot I_{n}^{[3]}(\mathbf{r},t)+{\Large b}%
_{\tau o}n(\mathbf{r},t)\mathbf{1}^{\left[  2\right]  }{\large +}%
S_{n}^{\left[  2\right]  }\left(  \mathbf{r},t\right)  , \label{eq70}%
\end{equation}

\[
\frac{\partial}{\partial t}{\Large I}_{n}^{\left[  3\right]  }%
{\large (\mathbf{r},t)+\nabla\cdot I}_{n}^{[4]}{\large (\mathbf{r}%
,t)}{\large =}%
\]%
\[
\frac{1}{m}\sum_{{\large s=1}}^{{\large 3}}{\huge \wp}(1,s)\left[
\boldsymbol{%
\mathcal{F}%
}\left(  \mathbf{r},t\right)  I_{n}^{[2]}(\mathbf{r},t)\right]
\]%
\[
-\theta_{3}^{-1}\text{ }{\Large I}_{n}^{\left[  3\right]  }{\large (\mathbf{r}%
,t)}%
\]%
\[
+{\Large a}_{LO}\sum_{{\large s=1}}^{{\large 3}}{\huge \wp}(1,s)\left[  \nabla
I_{n}^{[2]}(\mathbf{r},t)\right]
\]%
\[
+6\text{ }{\Large a}_{L1}\nabla\cdot I_{n}^{[4]}(\mathbf{r},t)
\]%
\begin{equation}
+{\Large b}_{\tau o}\left\{  \widehat{{\Huge \wp}}_{3}\text{ }\left[
\mathbf{1}^{\left[  2\right]  }\mathbf{I}_{n}{\large (\mathbf{r},t)}\right]
\right\}  +S_{n}^{\left[  3\right]  }\left(  \mathbf{r},t\right)  ,
\label{eq71}%
\end{equation}
where $\boldsymbol{%
\mathcal{F}%
}\left(  \mathbf{r},t\right)  $, given in Eq. (\ref{eqc1}), and the
expressions for $\mathbf{S}_{n}\left(  \mathbf{r},t\right)  $, $S_{n}^{\left[
2\right]  }\left(  \mathbf{r},t\right)  $ and $S_{n}^{\left[  3\right]
}\left(  \mathbf{r},t\right)  $ are given in Appendix C. The Maxwell times
$\theta_{1}^{-1}$, $\theta_{2}^{-1}$ and $\theta_{3}^{-1}$ are obtained from
Eq. (\ref{eq61}) respectively for $l=1,2$ and $3$. Moreover, as noticed, if we
multiply Eq. (\ref{eq70}) by the mass $m$, we do have an equation for the
pressure field tensor%
\begin{equation}
{\large P}^{[2]}{\large (\mathbf{r},t)=mI}_{n}^{[2]}{\large (\mathbf{r},t),}
\label{eq73}%
\end{equation}
composed of the hydrostatic contribution (the diagonal terms) and the shear
stress (the non-diagonal terms) and the convective pressure (cf. Eq.
(\ref{eq90}) presented later on, but where the shear contributions have been
discarded). We also mention that taking into account Eq. (\ref{eq87}) below
relating $\mathbf{I}_{n}{\large (\mathbf{r},t)}$ with the barycentric velocity
$\mathbf{v}\left(  \mathbf{r},t\right)  $, Eq. (\ref{eq69}) can be transformed
in an evolution equation for the latter to obtain a generalized Navier-Stokes
equation (future publication).

Let us now, for illustration, consider the case when we can perform \ a
truncation in a third order, that is, to consider as basic variables
$n(\mathbf{r},t)$ its flux $\mathbf{I}_{n}(\mathbf{r},t)$ and and the pressure
tensor ${\large mI}_{n}^{[2]}{\large (\mathbf{r},t)}$. In this contracted
description we consider Eqs. (\ref{eq68}), (\ref{eq69}) and (\ref{eq70}) but
with the further restrictions in Eqs. (\ref{eq69}) and (\ref{eq70}) of
neglecting: \textbf{1}. the shear stress contribution, more precisely
introducing the trace of the pressure tensor which is proportional to the
energy density $h(\mathbf{r},t)$, that is,
\begin{equation}
Tr\left\{  {\large P}^{[2]}{\large (\mathbf{r},t)}\right\}  =2h(\mathbf{r},t)
\label{eq74}%
\end{equation}
where convective pressure has been disconsidered, cf. Eq. (\ref{eq88}),
\textbf{2}. the terms with coefficients ${\Large a}_{L}$ whose origin is in
self-energy corrections, which simply would renormalize the kinetic
coefficients, and \textbf{3}. the terms $\mathbf{S}_{n}\left(  \mathbf{r}%
,t\right)  $ and $S_{n}^{\left[  2\right]  }\left(  \mathbf{r},t\right)  $
which contain the energy density $h(\mathbf{r},t)$ and its flux $\mathbf{I}%
_{h}(\mathbf{r},t)$ thus, disregarding thermo-striction effects.

The evolution equations for the chosen hydrodynamic variables, $n(\mathbf{r}%
,t)$, $\mathbf{I}_{n}(\mathbf{r},t)$ and ${\large I}_{n}^{[2]}%
{\large (\mathbf{r},t)}$, in the conditions above stated take the form%
\begin{equation}
\frac{\partial}{\partial t}n(\mathbf{r},t)+\nabla\cdot\mathbf{I}%
_{n}(\mathbf{r},t)=0, \label{eq75}%
\end{equation}

\bigskip%

\[
\frac{\partial}{\partial t}\mathbf{I}_{n}{\large (\mathbf{r},t)+\nabla\cdot
I}_{n}^{[2]}{\large (\mathbf{r},t)}{\large =}\frac{n(\mathbf{r},t)}%
{m}\boldsymbol{%
\mathcal{F}%
}\left(  \mathbf{r},t\right)
\]%
\begin{equation}
-\theta_{1}^{-1}\mathbf{I}_{n}{\large (\mathbf{r},t),} \label{eq76}%
\end{equation}

\bigskip%

\[
\frac{\partial}{\partial t}{\Large I}_{n}^{\left[  2\right]  }%
{\large (\mathbf{r},t)+\nabla\cdot I}_{n}^{[3]}{\large (\mathbf{r}%
,t)}{\large =}{\Large b}_{\tau o}n(\mathbf{r},t)\mathbf{1}^{\left[  2\right]
}%
\]%
\[
-\theta_{2}^{-1}{\large I}_{n}^{[2]}{\large (\mathbf{r},t)}%
\]%
\begin{equation}
{\large +}\frac{1}{m}\left\{  \left[  \boldsymbol{%
\mathcal{F}%
}\left(  \mathbf{r},t\right)  \mathbf{I}_{n}{\large (\mathbf{r},t)]+}\text{
}[\mathbf{I}_{n}{\large (\mathbf{r},t)}\text{ }\boldsymbol{%
\mathcal{F}%
}\left(  \mathbf{r},t\right)  \right]  \right\}  . \label{eq77}%
\end{equation}

Deriving in time Eq. (\ref{eq75}) and, next, in the result inserting
$\partial\mathbf{I}_{n}{\large (\mathbf{r},t)/\partial t}$, using Eq.
(\ref{eq76}) we arrive at the hyperbolic (Maxwell-Cattaneo type) evolution
equation for $n(\mathbf{r},t)$
\[
\frac{\partial^{2}}{\partial t^{2}}n(\mathbf{r},t)+\frac{1}{\theta_{1}}%
\frac{\partial}{\partial t}n(\mathbf{r},t)=\mathbf{\nabla}\cdot{\large \nabla
\cdot I}_{n}^{[2]}{\large (\mathbf{r},t)}%
\]%
\begin{equation}
-{\large \nabla\cdot}\left[  \frac{n(\mathbf{r},t)}{m}\boldsymbol{%
\mathcal{F}%
}\left(  \mathbf{r},t\right)  \right]  \label{eq78}%
\end{equation}
Furthermore, deriving in time this Eq. (\ref{eq78}) and using Eq. (\ref{eq77})
it follows that
\[
\frac{\partial^{3}}{\partial t^{3}}n(\mathbf{r},t)+\left[  \frac{1}{\theta
_{1}}+\frac{1}{\theta_{2}}\right]  \frac{\partial^{2}}{\partial t^{2}%
}n(\mathbf{r},t)
\]%
\[
+\frac{1}{\theta_{1}\theta_{2}}\frac{\partial}{\partial t}n(\mathbf{r}%
,t)={\Large b}_{\tau o}\nabla^{2}n(\mathbf{r},t)
\]%
\[
-\mathbf{\nabla}\cdot{\large \nabla\cdot\mathbf{\nabla}\cdot I}_{n}%
^{[3]}{\large (\mathbf{r},t)-}\frac{1}{\theta_{2}}{\large \nabla\cdot}\left[
\frac{n(\mathbf{r},t)}{m}\boldsymbol{%
\mathcal{F}%
}\left(  \mathbf{r},t\right)  \right]
\]%
\[
+\frac{1}{m}\nabla\cdot\nabla\cdot\left\{  \left[  \boldsymbol{%
\mathcal{F}%
}\left(  \mathbf{r},t\right)  \mathbf{I}_{n}{\large (\mathbf{r},t)]+}\text{
}[\mathbf{I}_{n}{\large (\mathbf{r},t)}\boldsymbol{%
\mathcal{F}%
}\left(  \mathbf{r},t\right)  \right]  \right\}
\]%
\begin{equation}
-\frac{\partial}{\partial t}{\large \nabla\cdot}\left[  \frac{n(\mathbf{r}%
,t)}{m}\boldsymbol{%
\mathcal{F}%
}\left(  \mathbf{r},t\right)  \right]  . \label{eq79}%
\end{equation}
\ The divergence of the third-order flux, ${\large \mathbf{\nabla}\cdot I}%
_{n}^{[3]}$, in terms of the basic variables is evaluated on the basis of
Eq.(\ref{eq77}), the evolution equation for the second-order flux, which we
recall, related to the pressure tensor [cf. Eq.(\ref{eq73})]. For that purpose
we consider conditions such that the pressure is changing slowly in time
($\theta_{2}\partial{\Large I}_{n}^{\left[  2\right]  }/\partial
t<<{\large I}_{n}^{[2]}$ or $\omega$ $\theta_{2}<<1$ along the motion), and
then from Eq.(\ref{eq77}) follows that
\[
{\large \mathbf{\nabla}\cdot I}_{n}^{[3]}{\large (\mathbf{r},t)=}%
{\Large b}_{\tau o}n(\mathbf{r},t)\mathbf{1}^{\left[  2\right]  }%
\]%
\[
-\text{ }\theta_{2}^{-1}{\large I}_{n}^{[2]}{\large (\mathbf{r},t)}%
\]%
\begin{equation}
{\large +}\frac{1}{m}\left\{  \left[  \boldsymbol{%
\mathcal{F}%
}\left(  \mathbf{r},t\right)  \mathbf{I}_{n}{\large (\mathbf{r},t)]+}\text{
}[\mathbf{I}_{n}{\large (\mathbf{r},t)}\boldsymbol{%
\mathcal{F}%
}\left(  \mathbf{r},t\right)  \right]  \right\}  , \label{eq80}%
\end{equation}
and using this result in Eq. (\ref{eq79}) it follows the equation%
\[
\frac{\partial^{3}}{\partial t^{3}}n(\mathbf{r},t)+\left[  \frac{1}{\theta
_{1}}+\frac{1}{\theta_{2}}\right]  \frac{\partial^{2}}{\partial t^{2}%
}n(\mathbf{r},t)
\]%
\[
+\frac{1}{\theta_{1}\theta_{2}}\frac{\partial}{\partial t}n(\mathbf{r}%
,t)=\frac{1}{\theta_{2}}{\large \nabla\cdot\mathbf{\nabla}\cdot I}_{n}%
^{[2]}{\large (\mathbf{r},t)}%
\]%
\begin{equation}
{\large -}\frac{1}{\theta_{2}}{\large \nabla\cdot}\left[  \frac{n(\mathbf{r}%
,t)}{m}\boldsymbol{%
\mathcal{F}%
}\left(  \mathbf{r},t\right)  \right]  -\frac{\partial}{\partial
t}{\large \nabla\cdot}\left[  \frac{n(\mathbf{r},t)}{m}\boldsymbol{%
\mathcal{F}%
}\left(  \mathbf{r},t\right)  \right]  . \label{eq81}%
\end{equation}

To close this Eq. (\ref{eq81}) it is necessary to evaluate $I_{n}^{[2]}$
which, we recall, is given by
\begin{equation}
I_{n}^{[2]}(\mathbf{r},t)=\int d^{3}p\text{ }\left[  \frac{\mathbf{p}}{m}%
\frac{\mathbf{p}}{m}\right]  f_{1}\left(  \mathbf{r},\mathbf{p};t\right)  .
\label{eq82}%
\end{equation}
We resort now to the use of Eq. (\ref{eq17}) and for $F_{1}(\mathbf{r}%
,\mathbf{p},t)$ we use an expansion in variable $\mathbf{p}$, namely%
\[
F_{1}(\mathbf{r},\mathbf{p},t)=F_{1n}(\mathbf{r},t)+\left.  \frac{\partial
F_{1}(\mathbf{r},\mathbf{p},t)}{\partial\mathbf{p}}\right\vert _{0}%
\cdot\mathbf{p}%
\]%
\[
+\frac{1}{2}\left.  \frac{\partial^{2}F_{1}(\mathbf{r},\mathbf{p},t)}{\partial
p^{2}}\right\vert _{0}\frac{p^{2}}{2m}%
\]%
\begin{equation}
\mathbf{+}\frac{1}{2}\left.  \frac{\partial^{2}F_{1}(\mathbf{r},\mathbf{p}%
,t)}{\partial\mathbf{p\partial p}}\right\vert _{0}\odot\left[  \overset{\circ
}{\mathbf{pp}}\right]  +..., \label{eq83}%
\end{equation}
where lower index nought indicates that the derivative is taken at
$\mathbf{p=0}$, and $\left[  \overset{\circ}{\mathbf{pp}}\right]  $ is the
traceless part of the tensor. We rewrite $F_{1}$ in the form
\[
F_{1}(\mathbf{r},\mathbf{p},t)\simeq\varphi_{n}(\mathbf{r},t)+\mathbf{F}%
_{n}(\mathbf{r},t)\cdot\mathbf{p/}m
\]%
\begin{equation}
+F_{h}(\mathbf{r},t)\frac{p^{2}}{2m} \label{eq84}%
\end{equation}
that is, keeping terms up to second order in $\mathbf{p}$. This is consistent
with the contracted description we used, and of disregarding the shear stress,
and where $\varphi_{n}$, $\mathbf{F}_{n}$ and $F_{h}$ are the nonequilibrium
thermodynamic variables conjugated to the density, the flux and the trace of
$I_{n}^{[2]}$ which is proportional to the energy. Moreover, introducing the
alternative forms%

\begin{equation}
\mathbf{F}_{n}(\mathbf{r},t)\equiv-m\beta(\mathbf{r},t)\mathbf{v}\left(
\mathbf{r},t\right)  , \label{eq85}%
\end{equation}%
\begin{equation}
F_{h}(\mathbf{r},t)\equiv\beta(\mathbf{r},t), \label{eq86}%
\end{equation}
it follows that%
\begin{equation}
\mathbf{I}_{n}(\mathbf{r},t)=n\left(  \mathbf{r},t\right)  \mathbf{v}\left(
\mathbf{r},t\right)  , \label{eq87}%
\end{equation}
defining the barycentric velocity $\mathbf{v}\left(  \mathbf{r},t\right)  .$
From Eqs. (\ref{eq28}), (\ref{eq29}) and (\ref{eq31}) there follows that
\[
h\left(  \mathbf{r},t\right)  =\frac{m}{2}Tr\left\{  I_{n}^{[2]}%
(\mathbf{r},t)\right\}  =\frac{3}{2}n\left(  \mathbf{r},t\right)  \beta
^{-1}(\mathbf{r},t)
\]%
\begin{equation}
+\frac{m}{2}n\left(  \mathbf{r},t\right)  \mathrm{v}^{2}(\mathbf{r},t),
\label{eq88}%
\end{equation}
where we can write $\beta^{-1}(\mathbf{r},t)=k_{B}T^{\ast}(\mathbf{r},t)$
introducing a nonequilibrium temperature (called quasitemperature
\cite{luzzi4}, \cite{luzzi5}), as well as
\[
\mathbf{I}_{h}(\mathbf{r},t)=\frac{5}{2}n\left(  \mathbf{r},t\right)
\beta^{-1}(\mathbf{r},t)\mathbf{v}\left(  \mathbf{r},t\right)
\]%
\begin{equation}
+\frac{m}{2}n\left(  \mathbf{r},t\right)  \mathrm{v}^{2}(\mathbf{r}%
,t)\mathbf{v}\left(  \mathbf{r},t\right)  , \label{eq89}%
\end{equation}
and
\[
{\large I}_{n}^{[2]}{\large (\mathbf{r},t)=}\frac{1}{m}\beta^{-1}%
(\mathbf{r},t)n\left(  \mathbf{r},t\right)  \mathbf{1}^{\left[  2\right]  }%
\]%
\begin{equation}
+n\left(  \mathbf{r},t\right)  \mathbf{v}\left(  \mathbf{r},t\right)
\mathbf{v}\left(  \mathbf{r},t\right)  . \label{eq90}%
\end{equation}
Introducing Eq. (\ref{eq90}) in Eq. (\ref{eq81}), if $\ {\large \nabla\cdot
}\left[  \frac{n(\mathbf{r},t)}{m}\boldsymbol{%
\mathcal{F}%
}\left(  \mathbf{r},t\right)  \right]  >>$ $\theta_{2}\frac{\partial}{\partial
t}{\large \nabla\cdot}\left[  \frac{n(\mathbf{r},t)}{m}\boldsymbol{%
\mathcal{F}%
}\left(  \mathbf{r},t\right)  \right]  $ we finally arrive at
\[
\theta_{1}\theta_{2}\frac{\partial^{3}}{\partial t^{3}}n(\mathbf{r},t)+\left[
\theta_{1}+\theta_{2}\right]  \frac{\partial^{2}}{\partial t^{2}}%
n(\mathbf{r},t)
\]%
\begin{equation}
+\frac{\partial}{\partial t}n(\mathbf{r},t)=-\mathbf{\nabla}\cdot
\mathbf{j}_{n}(\mathbf{r},t), \label{eq92}%
\end{equation}
where%
\begin{equation}
\mathbf{j}_{n}(\mathbf{r},t)=-\mathcal{D}^{\left[  2\right]  }(\mathbf{r}%
,t)\cdot\nabla n\left(  \mathbf{r},t\right)  -n\left(  \mathbf{r},t\right)
\boldsymbol{V}\left(  \mathbf{r},t\right)  \label{eq93}%
\end{equation}
plays the role of a generalized flux with at the right being present a
generalized thermodynamic force, and where
\begin{equation}
\mathcal{D}^{\left[  2\right]  }(\mathbf{r},t)=\theta_{1}\left[  \frac
{k_{B}T^{\ast}(\mathbf{r},t)}{m}\mathbf{1}^{\left[  2\right]  }+\left[
\mathbf{v}\left(  \mathbf{r},t\right)  \mathbf{v}\left(  \mathbf{r},t\right)
\right]  \right]  \label{eq94}%
\end{equation}
is playing the role of a generalized diffusion tensor (composed of two parts,
a first one of thermal origin and a second associated to the drift of the
material) and%

\[
\boldsymbol{V}\left(  \mathbf{r},t\right)  =\theta_{1}\left[  \mathbf{\nabla
}\left(  \frac{k_{B}T^{\ast}(\mathbf{r},t)}{m}\right)  \right]
\]%
\begin{equation}
+\theta_{1}\left[  \mathbf{\nabla\cdot\lbrack vv]}-\frac{1}{m}\boldsymbol{%
\mathcal{F}%
}\left(  \mathbf{r},t\right)  \right]  , \label{eq95}%
\end{equation}
composed of three terms, one of thermal origin another coming from the
drifting movement and a third from the applied force.

Moreover, in the steady state ($\partial n/\partial t=0$ and then
$\mathbf{j}=0$) there follows that the density satisfies the equation%
\begin{equation}
\mathcal{D}^{\left[  2\right]  }(\mathbf{r})\cdot\nabla n\left(
\mathbf{r}\right)  =-n\left(  \mathbf{r}\right)  \boldsymbol{V}\left(
\mathbf{r}\right)  . \label{eq96}%
\end{equation}

Returning to Eq.(\ref{eq92}), its Fourier transform reads
\begin{align}
&  i\omega\left[  -\omega^{2}\theta_{1}\theta_{2}-i\omega\left(  \theta
_{1}+\theta_{2}\right)  +1\right]  n(\mathbf{Q},\omega)\nonumber\\
&  =i\mathbf{Q}\cdot\mathbf{j}_{n}(\mathbf{Q},\omega), \label{eq961}%
\end{align}
which give us an illustration on the criterion of contraction of description:
1. In conditions such that $\omega^{2}\theta_{1}\theta_{2}<<1$, the term with
third time derivative can be neglected and the evolution equation acquires the
form of a generalized hyperbolic Maxwell-Cattaneo one. 2. If further
$\omega\left(  \theta_{1}+\theta_{2}\right)  <<1$ the second time derivative
also can be neglected and we are left with a generalized parabolic
diffusion-like equation, and in that way there follows a chain of increasing
contractions of description of the hydrodynamic motion.

Finally, to perform numerical calculations and analyze the results we
introduce a central force interaction between particles in the system with
those in the bath of the Gaussian form, called the Gaussian core model (GCM)
\cite{stillinger},%
\begin{equation}
w\left(  r\right)  =\frac{U}{\sqrt{2\pi r_{o}^{2}}}\exp\left\{  -r^{2}%
/r_{o}^{2}\right\}  , \label{eq97}%
\end{equation}
with the open parameters $U$ and $r_{o}$ ( $r_{o}$ is a length scale playing
the role of a range length and $U/r_{o}$ being the interaction strength ). It
has been noticed that this kind of potential belongs to the class of
interactions which do not diverge at the origin, i.e., are bounded. They are
potentials corresponding to effective interactions between the centers of mass
of soft, flexible macromolecules such as polymer chains \cite{louis},
dendrimers \cite{likos}, and others. The centers of mass of two macromolecules
can coincide without violation of the excluded volume conditions, hence
implying in a bounded interaction \cite{likos1}. Several studies of this
potential can be consulted, for example, in Refs. \cite{stillinger1}-
\cite{stillinger4}.

It can be noticed that GCM of Eq.(\ref{eq97}) roughly mimics a hard sphere
potential with radius $r_{o}$, and that in the limit of $r_{o}$ going to zero
goes over a contact potential $U$ $\delta\left(  r\right)  .$ The Fourier
transform is%
\begin{equation}
\psi\left(  \mathbf{Q}\right)  =\frac{\pi}{\sqrt{2}}Ur_{o}^{2}\exp\left\{
-\frac{1}{4}r_{o}^{2}Q^{2}\right\}  , \label{eq98}%
\end{equation}
In terms of these results we find that
\begin{equation}
{\large \theta}_{1}^{-1}=\frac{\sqrt{\pi}}{6\sqrt{2}}\frac{n_{R}}{M^{1/2}%
}\frac{M}{m}\left(  1+\frac{M}{m}\right)  \beta_{o}^{3/2}U^{2}, \label{eq99}%
\end{equation}
what tells us that the momentum relaxation time becomes very large for the
Brownian particle when $m>>M$, and very small for the Lorentz particle when
$m<<M$. Furthermore, that ${\large \theta}_{1}$ increases with the power $3/2$
of the temperature $T_{o}$ and, as expected, with the reciprocal of the
density of scattering centers.

Moreover, it can also be noticed that in the limit of a contact potential
$\left(  r_{o}\rightarrow0\right)  $, the quantity
\begin{equation}
\frac{1}{\kappa}=\frac{\sqrt{\pi}}{2^{3/2}}\left[  1+\frac{m}{M}\right]
^{-1}r_{o} \label{eq100}%
\end{equation}
tells us that $\kappa^{-1}$ goes to zero and then [cf. Eqs. (\ref{eq114}) and
(\ref{eq118})] the kinetic coefficients ${\large a}_{{\Large Lo}}$ and
${\Large a}_{l1}$ approach zero, i. e. ${\large J}_{L}{\large (\mathbf{r},t)}$
does not contribute.

\section{Concluding Remarks}

Briefly summarizing the results, it has been shown how a statistical
nonequilibrium ensemble formalism (applicable to the study of systems even in
conditions far-from equilibrium) provides a microscopic foundation for a
Nonlinear Higher-Order Hydrodynamics. Its description is based on the set of
hydrodynamic variables consisting of the densities of energy and matter
(particles) and their fluxes of all order.

All these hydrodynamic variables are the average value over the nonequilibrium
ensemble of the corresponding microscopic mechanical operators. Once the
complete set of macrovariables is given we can obtain the nonlinear
hydrodynamic equations, which are the average value over the nonequilibrium
ensemble of Hamilton equations of motion (in the classical level or Heisenberg
equations at the quantum level) of the basic microvariables (mechanical
observables). Once all these hydrodynamical variables, cf. Eqs. (\ref{eq28})
to (\ref{eq32}) and (\ref{eq35}) to (\ref{eq37}), involve the single particle
distribution function $f_{1}\left(  \mathbf{r},\mathbf{p};t\right)  $, their
evolution equations follow from the evolution equation for $f_{1}$ [cf. Eq.
(\ref{eq42})]. It is obtained the set of evolution equations given in Eq.
(\ref{eq43}): $l=0$ for the density, $l=1$ for the first (vectorial) flux,
$l\geq2$ for the higher order tensorial fluxes, all of which are coupled together.

These generalized hydrodynamic equations present on the left side the
conserving part of the corresponding quantity, and on the right-hand side are
present the collision integrals which include the action of external sources
and the contributions of scattering processes responsible for dissipative effects.

In that way we do have a quite generalized hydrodynamics under any arbitrary
condition of excitation, which, as noticed, can be referred to as
\textit{Mesoscopic Hydro-Thermodynamics.}

\begin{acknowledgments}
We acknowledge financial support from S\~{a}o Paulo State Research Foundation
\ (FAPESP). ARV\ and RL are Brazil National Research Council (CNPq) research
fellows. CABS acknowledge a leave of absence granted by the Brazilian
Technological Institute of \ Aeronautics, and is grateful to the Condensed
Matter Physics Department at the University of Campinas for the kind
hospitality there received.
\end{acknowledgments}

\appendix{}

\section{\bigskip Tensorial Coefficients of Eqs. (\ref{eq50})-(\ref{eq53}) and
Last Term in Eq. (\ref{eq60})}

The kinetic tensorial coefficients of Eq. (\ref{eq50}) to Eq. (\ref{eq53}) are:%

\begin{equation}
{\large \Lambda}_{{\Large \tau o}}^{[2k+2]}=\sum_{\mathbf{Q}}{\large g}%
_{{\Large \tau k}}\left(  Q\right)  \mathbf{Q}^{\left[  2k+2\right]  },
\label{eqa1}%
\end{equation}

\begin{equation}
{\large \Lambda}_{{\Large \tau}}^{[2k+2]}=\sum_{\mathbf{Q}}{\large f}%
_{{\Large \tau k}}\left(  Q\right)  \mathbf{Q}^{\left[  2k+2\right]  },
\end{equation}

\begin{equation}
{\large \Lambda}_{{\Large L}}^{[2k+2]}=\sum_{\mathbf{Q}}{\large f}%
_{{\Large Lk}}\left(  Q\right)  \mathbf{Q}^{\left[  2k+2\right]  },
\end{equation}%
\begin{equation}
{\large \Lambda}_{NL1}^{{\large [2k+1]}}\left(  \mathbf{r}^{\prime}%
-\mathbf{r}\right)  =i\frac{n_{R}\text{ }\beta_{o}}{\mathcal{V}}%
\sum_{\mathbf{Q}}\left\vert \psi\left(  Q\right)  \right\vert ^{2}%
{\large e}^{{\large i\mathbf{Q}\cdot}\left(  \mathbf{r}^{\prime}%
\mathbf{-r}\right)  }\nonumber
\end{equation}%
\begin{equation}
\times\frac{\left(  -1\right)  ^{k}}{\left(  2k-1\right)  !!}\left(
\frac{M\beta_{o}}{Q^{2}}\right)  ^{k}\mathbf{Q}^{\left[  2k+1\right]  },
\end{equation}%
\begin{equation}
{\large \Lambda}_{NL2}^{{\large [2k+2]}}\left(  \mathbf{r}^{\prime}%
\mathbf{-r}\right)  =\frac{n_{R}\text{ }\beta_{o}}{\mathcal{V}}\frac{\left(
M\beta_{o}\right)  ^{1/2}\pi}{\sqrt{2\pi}\text{ }}\nonumber
\end{equation}%
\begin{equation}
\times\sum_{\mathbf{Q}}\frac{\left\vert \psi\left(  Q\right)  \right\vert
^{2}}{Q}{\large e}^{{\large i\mathbf{Q}\cdot}\left(  \mathbf{r}^{\prime
}\mathbf{-r}\right)  }\frac{\left(  -1\right)  ^{k}}{k!}\left(  \frac
{M\beta_{o}}{2Q^{2}}\right)  ^{k}\mathbf{Q}^{\left[  2k+2\right]  }.
\end{equation}
Where we have defined,%
\begin{equation}
{\large g}_{{\Large \tau k}}(Q)=\frac{n_{R}}{\mathcal{V}}\frac{\sqrt{2\pi
M\beta_{o}}}{m^{2}}\frac{\left\vert \psi\left(  Q\right)  \right\vert ^{2}}%
{Q}\frac{\left(  -1\right)  ^{k}}{k!}\left(  \frac{M\beta_{o}}{2Q^{2}}\right)
^{k},
\end{equation}%
\[
{\large f}_{{\Large \tau k}}\left(  Q\right)  =\frac{n_{R}}{\mathcal{V}}%
\frac{\left(  M\beta_{o}\right)  ^{3/2}\pi}{\sqrt{2\pi}\text{ }m}%
\frac{\left\vert \psi\left(  Q\right)  \right\vert ^{2}}{Q}\left(  \frac{1}%
{M}+\frac{1}{m}\right)
\]%
\begin{equation}
\frac{\left(  -1\right)  ^{k+1}}{k!}\left(  \frac{M\beta_{o}}{2Q^{2}}\right)
^{k}, \label{eqa7}%
\end{equation}%
\begin{equation}
{\large f}_{{\Large Lk}}\left(  Q\right)  =\frac{n_{R}}{\mathcal{V}}%
\frac{M\beta_{o}}{m^{2}}\frac{\left\vert \psi\left(  Q\right)  \right\vert
^{2}}{Q^{2}}\frac{\left(  -1\right)  ^{k+1}}{\left(  2k-1\right)  !!}\left(
\frac{M\beta_{o}}{Q^{2}}\right)  ^{k}. \label{eqa8}%
\end{equation}

The last term on the right of Eq. (\ref{eq60}) is given by,%

\begin{equation}
{\large R}_{n}^{\left[  l\right]  }{\large (\mathbf{r},t)=J}_{{\Large \tau}%
R}^{\left[  l\right]  }{\large (\mathbf{r},t)+J}_{LR}^{\left[  l\right]
}{\large (\mathbf{r},t)}\text{,}%
\end{equation}%
\[
{\large J}_{LR}^{\left[  l\right]  }{\large (}\mathbf{r}{\large ,t)=}%
\sum_{{\large k=2}}^{{\large \infty}}\sum_{{\large s=1}}^{{\large l}%
}{\huge \wp}{\large (}1{\large ,s)}\left[  \Lambda_{L}^{[2k+2]}\right.
\]%
\begin{equation}
\left.  \odot\nabla I_{n}^{[2k+l-1]}(\mathbf{r},t)\right]  \text{,}%
\end{equation}%
\[
{\large J}_{{\large \tau}R}^{\left[  l\right]  }{\large (}\mathbf{r}%
{\large ,t)=}\sum_{{\large k=2}}^{{\large \infty}}\sum_{{\large s=1}%
}^{{\large l}}{\huge \wp}{\large (}1{\large ,s)}\left[  \Lambda_{{\large \tau
}}^{[2k+2]}\odot I_{n}^{[2k+l]}(\mathbf{r},t)\right]
\]%
\begin{equation}
+\sum_{k{\large =2}}^{{\large \infty}}\left\{  \widehat{{\Huge \wp}}_{l}\text{
}\left[  \Lambda_{\tau o}^{[2k+2]}\odot I_{n}^{[2k+l-2]}(\mathbf{r},t)\right]
\right\}  \text{,}%
\end{equation}
where the operators ${\huge \wp}{\large (}1{\large ,s)}$ and
$\widehat{{\Huge \wp}}_{l}$ are defined in the main text.

\section{The Kinetic Coefficients in Eqs. (\ref{eq59}) and (\ref{eq60})}

We do have that,%
\begin{equation}
{\Large a}_{{\Large \tau o}}=\frac{\mathcal{V}}{\left(  2\pi\right)  ^{3}%
}\frac{4\pi}{3}\int dQ\text{ }Q^{4}f_{\tau o}\left(  Q\right)  \label{eqb1}%
\end{equation}
with $k=0$ in Eq. (\ref{eq7}). And%
\begin{equation}
{\large f}_{{\Large \tau o}}\left(  Q\right)  =-\frac{n_{R}}{\mathcal{V}}%
\frac{\left(  M\beta_{o}\right)  ^{3/2}\pi}{\sqrt{2\pi}m^{2}}\frac{\left\vert
\psi\left(  Q\right)  \right\vert ^{2}}{Q}\left(  \frac{m}{M}+1\right)
\end{equation}
where $\psi\left(  Q\right)  $ is the Fourier transform of the potential
energy $w\left(  \left\vert \mathbf{r}_{j}-\mathbf{R}_{\mu}\right\vert
\right)  $, $n_{R}$ is the density of particles in the thermal bath,
$\mathcal{V}$ is the volume, and $\beta_{o}^{-1}=k_{B}T_{o}$. Moreover,
\begin{equation}
{\Large a}_{{\Large \tau1}}=-\frac{M\beta_{o}}{10}{\Large a}_{{\Large \tau o}%
},
\end{equation}%
\begin{equation}
{\Large a}_{{\Large Lo}}=\sqrt{\frac{2}{M\beta_{o}\pi}}\frac{\mathrm{1}%
}{\kappa}{\Large a}_{{\Large \tau o}}, \label{eq114}%
\end{equation}%
\begin{equation}
\frac{1}{\kappa}=\frac{\int dQ\text{ }Q^{2}\left\vert \psi\left(  Q\right)
\right\vert ^{2}}{\int dQ\text{ }Q^{3}\left\vert \psi\left(  Q\right)
\right\vert ^{2}\left(  1+\frac{m}{M}\right)  },
\end{equation}%
\begin{equation}
{\Large b}_{{\Large \tau o}}=-\frac{2}{M\text{ }\beta_{o}\text{ }}%
{\Large a}_{{\Large \tau o}}\left(  1+\frac{m}{M}\right)  ^{-1},
\end{equation}%
\begin{equation}
{\Large b}_{{\Large \tau1}}=\frac{{\Large a}_{{\Large \tau o}}}{5}\left(
1+\frac{m}{M}\right)  ^{-1}, \label{eq117}%
\end{equation}%
\begin{equation}
{\Large a}_{L1}=-\frac{1}{5\kappa}\sqrt{\frac{2M\beta_{o}}{\pi}}%
{\Large a}_{\tau o}. \label{eq118}%
\end{equation}

\section{The Last Terms of Eqs. (\ref{eq69}), (\ref{eq70}) and (\ref{eq71})}

The contributions present in Eqs. (\ref{eq69}), (\ref{eq70}) and (\ref{eq71})
in section VI are%
\begin{equation}
\boldsymbol{%
\mathcal{F}%
}\left(  \mathbf{r},t\right)  =-\nabla V_{ext}\left(  \mathbf{r};t\right)
-\mathbf{F}_{NL}\left(  \mathbf{r};t\right)  , \label{eqc1}%
\end{equation}%
\[
\mathbf{S}_{n}\left(  \mathbf{r},t\right)  =3{\Large a}_{{\Large \tau1}}%
\frac{2}{m}\mathbf{I}_{h}{\large (\mathbf{r},t)+}\frac{2}{m}{\Large a}%
_{L1}\nabla h(\mathbf{r},t)
\]%
\begin{equation}
+\mathbf{R}_{n}{\large (\mathbf{r},t),}%
\end{equation}

\bigskip%

\[
S_{n}^{\left[  2\right]  }\left(  \mathbf{r},t\right)  =\frac{2}{m}%
{\Large b}_{{\Large \tau1}}h(\mathbf{r},t)\mathbf{1}^{\left[  2\right]
}+6{\Large a}_{{\Large \tau1}}\frac{2}{m}I_{h}^{[2]}(\mathbf{r},t)
\]%
\begin{equation}
+{\Large a}_{{\Large L1}}\frac{2}{m}\left\{  \nabla\mathbf{I}_{h}%
(\mathbf{r},t)+\left[  \nabla\mathbf{I}_{h}(\mathbf{r},t)\right]
^{tr}\right\}  +R_{n}^{\left[  2\right]  }{\large (\mathbf{r},t),}%
\end{equation}
where upper index $tr$ stands for transpose,

\bigskip%

\[
S_{n}^{\left[  3\right]  }\left(  \mathbf{r},t\right)  ={\Large b}%
_{{\Large \tau1}}\frac{2}{m}\left\{  \widehat{{\Huge \wp}}_{3}\text{ }\left[
\mathbf{1}^{\left[  2\right]  }\mathbf{I}_{h}(\mathbf{r},t)\right]  \right\}
\]%
\[
+{\Large a}_{{\Large L1}}\frac{2}{m}\sum_{{\large s=1}}^{{\large 3}}%
{\huge \wp}{\large (}1{\large ,s)}\left[  \nabla I_{h}^{[2]}(\mathbf{r}%
,t)\right]
\]%
\begin{equation}
+9{\Large a}_{{\Large \tau1}}\frac{2}{m}I_{h}^{[3]}(\mathbf{r},t)+R_{n}%
^{\left[  3\right]  }{\large (\mathbf{r},t).}%
\end{equation}
In which,
\[
\mathbf{R}_{n}{\large (\mathbf{r},t)=}%
{\textstyle\sum\limits_{k=2}^{\infty}}
\left\{  {\large \Lambda}_{{\Large \tau}}^{[2k+2]}\odot I_{n}^{[2k+1]}%
(\mathbf{r},t)\right.
\]%
\begin{equation}
\left.  +{\large \Lambda}_{{\Large L}}^{[2k+2]}\odot\nabla I_{n}%
^{[2k]}(\mathbf{r},t)\right\}  ,
\end{equation}

\bigskip%

\[
R_{n}^{\left[  2\right]  }{\large (\mathbf{r},t)=}%
{\textstyle\sum\limits_{k=2}^{\infty}}
\{{\large \Lambda}_{{\Large \tau o}}^{[2k+2]}\odot I_{n}^{[2k]}(\mathbf{r},t)
\]%
\[
+{\large \Lambda}_{{\Large \tau}}^{[2k+2]}\odot I_{n}^{[2k+2]}(\mathbf{r},t)
\]%
\[
+I_{n}^{[2k+2]}(\mathbf{r},t)\odot{\large \Lambda}_{{\Large \tau}}%
^{[2k+2]}+{\large \Lambda}_{{\Large L}}^{[2k+2]}\odot\nabla I_{n}%
^{[2k+1]}(\mathbf{r},t)
\]%
\begin{equation}
+\left[  {\large \Lambda}_{{\Large L}}^{[2k+2]}\odot\nabla I_{n}%
^{[2k+1]}(\mathbf{r},t)\right]  ^{tr}\},
\end{equation}

\bigskip%

\begin{equation}
R_{n}^{\left[  3\right]  }{\large (\mathbf{r},t)=J}_{{\Large \tau}R}^{\left[
3\right]  }{\large (\mathbf{r},t)+J}_{LR}^{\left[  3\right]  }%
{\large (\mathbf{r},t),}%
\end{equation}

\bigskip%

\[
{\normalsize J}_{LR}^{\left[  3\right]  }{\normalsize (r,t)=}%
\]%
\begin{equation}
=\sum_{k{\large =2}}^{{\large \infty}}\sum_{{\large s=1}}^{{\large 3}%
}{\normalsize \wp(1,s)}\left[  {\large \Lambda}_{{\Large L}}^{[2k+2]}%
\odot\nabla I_{n}^{[2k+2]}(\mathbf{r},t)\right]  {\normalsize ,}%
\end{equation}

\bigskip%

\[
{\large J}_{{\Large \tau}R}^{\left[  3\right]  }{\large (\mathbf{r},t)=}%
\sum_{k{\large =2}}^{{\large \infty}}\widehat{{\Huge \wp}}_{3}\text{ }\left[
{\large \Lambda}_{{\Large \tau o}}^{[2k+2]}\odot I_{n}^{[2k+1]}(\mathbf{r}%
,t)\right]  ,
\]%
\begin{equation}
+\sum_{k{\large =2}}^{{\large \infty}}\sum_{{\large s=1}}^{{\large 3}%
}{\huge \wp}{\large (}1{\large ,s)}\left[  {\large \Lambda}_{{\Large \tau}%
}^{[2k+2]}\odot I_{n}^{[2k+3]}(\mathbf{r},t)\right]  .
\end{equation}

The several kinetic tensorial coefficients ${\large \Lambda}^{[r]}$ are given
in Appendix A.

\end{document}